\documentclass[runningheads]{svmult}

\usepackage{epsfig}

\usepackage{epsfig}

\usepackage{axodraw}

\usepackage{rotate}

\usepackage{makeidx}   
\usepackage{graphicx}  
\usepackage{subeqnar}  
\usepackage{multicol}  
\usepackage{physprbb}  
\makeindex             

\newcommand{\lsim}{
\mathrel{\hbox{\rlap{\hbox{\lower4pt\hbox{$\sim$}}}\hbox{$<$}}}}
\newcommand{\gsim}{
\mathrel{\hbox{\rlap{\hbox{\lower4pt\hbox{$\sim$}}}\hbox{$>$}}}}

\begin{document}


\begin{titlepage}
\begin{flushright}
\begin{tabular}{l}
CERN-TH/2002-293\\
hep--ph/0210323\\
October 2002
\end{tabular}
\end{flushright}

\vspace*{1.3truecm}

\begin{center}
\boldmath
{\Large \bf B Physics and CP Violation}
\unboldmath

\vspace*{1.6cm}

\smallskip
\begin{center}
{\sc {\large Robert Fleischer}}\\
\vspace*{2mm}
{\sl Theory Division, CERN, CH-1211 Geneva 23, Switzerland}
\end{center}

\vspace{2.0truecm}

{\large\bf Abstract\\[10pt]} \parbox[t]{\textwidth}{
After an introduction to the Standard-Model description of CP violation,
we turn to the main focus of these lectures, the $B$-meson system. Since 
non-leptonic $B$ decays play the key r\^ole for the exploration of CP 
violation, we have to discuss the tools to describe these transitions 
theoretically before classifying the main strategies to study CP violation. 
We will then have a closer look at the $B$-factory benchmark modes 
$B_d\to J/\psi K_{\rm S}$, $B_d\to\phi K_{\rm S}$ and $B_d\to\pi^+\pi^-$, 
and shall emphasize the importance of studies of $B_s$ decays at hadron
colliders. Finally, we focus on more recent developments related to 
$B\to\pi K$ modes and the $B_d\to\pi^+\pi^-$, $B_s\to K^+K^-$ system. 
}

\vspace{1.5cm}
 
{\sl Invited lecture at the International School ``Heavy Quark Physics'',\\
Dubna, Russia, 27 May -- 5 June 2002\\
To appear in the Proceedings (Lecture Notes in Physics)}
\end{center}

\end{titlepage}
 
\thispagestyle{empty}
\vbox{}
\newpage
 
\setcounter{page}{1}
 

%
\title*{B Physics and CP Violation}
%
%
%
%
\titlerunning{B Physics and CP Violation}
%
\author{Robert Fleischer}
%
%
%
\institute{Theory Division, CERN, CH-1211 Geneva 23, Switzerland}

\maketitle              

\begin{abstract}
After an introduction to the Standard-Model description of CP violation,
we turn to the main focus of these lectures, the $B$-meson system. Since 
non-leptonic $B$ decays play the key r\^ole for the exploration of CP 
violation, we have to discuss the tools to describe these transitions 
theoretically before classifying the main strategies to study CP violation. 
We will then have a closer look at the $B$-factory benchmark modes 
$B_d\to J/\psi K_{\rm S}$, $B_d\to\phi K_{\rm S}$ and $B_d\to\pi^+\pi^-$, 
and shall emphasize the importance of studies of $B_s$ decays at hadron
colliders. Finally, we focus on more recent developments related to 
$B\to\pi K$ modes and the $B_d\to\pi^+\pi^-$, $B_s\to K^+K^-$ system.  
\end{abstract}

\section{Introduction}\label{sec:intro}
The non-conservation of the CP symmetry, where C and P denote the 
charge-conjugation and parity transformation operators, respectively, 
is one of the most exciting phenomena in particle physics since
its unexpected discovery through $K_{\rm L}\to \pi^+\pi^-$ decays in 
1964 \cite{CP-discovery}. At that time it was believed that --
although weak interactions are neither invariant under P, nor invariant 
under C -- the product CP was preserved. Consider, for instance, the process
\begin{equation}
\pi^+\to e^+\nu_e~\stackrel{C}{\longrightarrow}~\pi^-\to 
e^-\nu_e^C~\stackrel{P}{\longrightarrow}~\pi^-\to e^-\overline{\nu}_e.
\end{equation}
Here the left-handed $\nu_e^C$ state is not observed in nature; only after
performing an additional P transformation do we obtain the right-handed
electron antineutrino. 

Before the start of the $B$ factories, CP-violating effects could only
be studied in the kaon system, where we distinguish between ``indirect''
CP violation, which is due to the fact that the mass eigenstates 
$K_{\rm S}$ and $K_{\rm L}$ of the neutral kaon system are not eigenstates 
of the CP operator, and ``direct'' CP violation, arising directly at
the decay amplitude level of the neutral kaon system. The former kind
of CP violation was already discovered in 1964 and is described by a
complex parameter $\varepsilon$, whereas the latter one, described
by the famous parameter $\mbox{Re}(\varepsilon'/\varepsilon)$, could 
only be established in 1999 after tremendous efforts by the NA48 (CERN) 
\cite{NA48} and KTeV (Fermilab) \cite{KTEV} collaborations, reporting
the following results in 2002:
\begin{equation}
\mbox{Re}(\varepsilon'/\varepsilon)=\left\{
\begin{array}{ll}
(14.7\pm2.2)\times10^{-4} & \mbox{(NA48 \cite{NA48-02})}\\
(20.7\pm2.8)\times10^{-4} & \mbox{(KTeV \cite{KTEV-02}).}
\end{array}\right.
\end{equation}
Unfortunately, the theoretical interpretation of 
$\mbox{Re}(\varepsilon'/\varepsilon)$ is still affected by large
hadronic uncertainties and does not provide a stringent test of the
Standard-Model description of CP violation, unless significant theoretical 
progress concerning the relevant hadronic matrix elements can be made 
\cite{bertolini,buras-KAON,CiMa}. 

In 2001, CP violation could also be established in $B$-meson decays by 
the BaBar (SLAC) \cite{BaBar-CP-obs} and Belle (KEK) \cite{Belle-CP-obs} 
collaborations, representing the start of a new era in the exploration
of CP violation. As we will discuss in these lecture notes, decays of 
neutral and charged $B$-mesons provide valuable insights into this 
phenomenon, offering in particular powerful tests of the Kobayashi--Maskawa 
(KM) mechanism \cite{KM}, which allows us to accommodate CP violation in 
the Standard Model of electroweak interactions. In Section~\ref{sec:SM},
we shall have a closer look at the Standard-Model description of
CP violation, and shall introduce the Wolfenstein parametrization and the
unitarity triangles of the Cabibbo--Kobayashi--Maskawa (CKM) matrix.  
Since non-leptonic decays of $B$ mesons play the key r\^ole in the 
exploration of CP violation, we have to discuss the tools to deal with 
these transitions and the corresponding theoretical problems in 
Section~\ref{sec:non-lept}. The main strategies to study CP violation are 
then classified in Section~\ref{sec:cp-strat}, before we focus on benchmark 
modes for the $B$ factories in Section~\ref{sec:b-fact-bench}. The great 
physics potential of $B_s$-meson decays for experiments at hadron colliders 
is emphasized in Section~\ref{sec:Bs-bench}, and will also be employed in 
Section~\ref{sec:recent}, where we discuss interesting recent developments.
Finally, we make a few comments on the ``usual'' rare $B$ decays in 
Section~\ref{sec:rare}, before we summarize our conclusions and give a 
brief outlook in Section~\ref{sec:concl}. 

A considerably more detailed presentation of CP violation in the
$B$ system can be found in \cite{RF-Phys-Rep}, as well as in the 
textbooks listed in \cite{CP-Books}. Another lecture on related 
topics was given by Neubert at this school \cite{neubert}.

\section{CP Violation in the Standard Model}\label{sec:SM}
\subsection{Charged-Current Interactions of Quarks}
The CP-violating effects discussed in these lectures originate from
the charged-current interactions of the quarks, described by the 
Lagrangian
\begin{equation}\label{cc-lag2}
{\cal L}_{\mbox{{\scriptsize int}}}^{\mbox{{\scriptsize CC}}}=
-\frac{g_2}{\sqrt{2}}\left(\begin{array}{ccc}\bar
u_{\mbox{{\scriptsize L}}},& \bar c_{\mbox{{\scriptsize L}}},&
\bar t_{\mbox{{\scriptsize L}}}\end{array}\right)\gamma^\mu\,\hat
V_{\mbox{{\scriptsize CKM}}}
\left(
\begin{array}{c}
d_{\mbox{{\scriptsize L}}}\\
s_{\mbox{{\scriptsize L}}}\\
b_{\mbox{{\scriptsize L}}}
\end{array}\right)W_\mu^\dagger \, + \, \mbox{h.c.,}
\end{equation}
where the gauge coupling $g_2$ is related to the gauge group 
$SU(2)_{\mbox{{\scriptsize L}}}$, the $W_\mu^{(\dagger)}$ field 
corresponds to the charged $W$ bosons, and $\hat
V_{\mbox{{\scriptsize CKM}}}$ denotes the CKM matrix, connecting the 
electroweak eigenstates of the down, strange and bottom quarks with 
their mass eigenstates through a unitary transformation. 

Since the CKM matrix elements $V_{UD}$ and $V_{UD}^\ast$ enter in 
$D\to U W^-$ and the CP-conjugate process $\overline{D}\to\overline{U}W^+$, 
respectively, where $D\in\{d,s,b\}$ and $U\in\{u,c,t\}$, we observe 
that the phase structure of the CKM matrix is closely related to CP 
violation. It was pointed out by Kobayashi and 
Maskawa in 1973 that actually one complex phase is required -- in addition to 
three generalized Euler angles -- to parametrize the quark-mixing 
matrix in the case of three fermion generations, thereby allowing us to 
accommodate CP violation in the Standard Model \cite{KM}. 

More detailed investigations show that additional conditions have to be 
satisfied for CP violation. They can be summarized as follows:
\begin{eqnarray}
\lefteqn{(m_t^2-m_c^2)(m_t^2-m_u^2)(m_c^2-m_u^2)}\nonumber\\
&&\times(m_b^2-m_s^2)(m_b^2-m_d^2)(m_s^2-m_d^2)\times
J_{\rm CP}\,\not=\,0,
\end{eqnarray}
where the Jarlskog parameter 
\begin{equation}
J_{\rm CP}=\pm\,\mbox{Im}\left(V_{i\alpha}V_{j\beta}V_{i\beta}^\ast 
V_{j\alpha}^\ast\right)\quad(i\not=j,\,\alpha\not=\beta)
\end{equation}
represents a measure of the ``strength'' of CP violation within the 
Standard Model \cite{jarlskog}. As data imply $J_{\rm CP}={\cal O}(10^{-5})$,
CP violation is a small effect in the Standard Model. In scenarios for
physics beyond the Standard Model, typically also new sources of CP
violation arise \cite{NP}.

\subsection{Wolfenstein Parametrization}
The quark transitions caused by charged-current interactions exhibit
an interesting hierarchy, which is made explicit in the Wolfenstein
parametrization of the CKM matrix \cite{wolf}:
\begin{equation}
\hat V_{\mbox{{\scriptsize CKM}}} =\left(\begin{array}{ccc}
1-\frac{1}{2}\lambda^2 & \lambda & A\lambda^3(\rho-i \eta) \\
-\lambda & 1-\frac{1}{2}\lambda^2 & A\lambda^2\\
A\lambda^3(1-\rho-i \eta) & -A\lambda^2 & 1
\end{array}\right)+{\cal O}(\lambda^4).
\end{equation}
This parametrization corresponds to an expansion in powers of the small 
quantity $\lambda=0.22$, which can be fixed through semileptonic kaon 
decays. The other parameters are of order 1, where $\eta$ leads to an
imaginary part of the CKM matrix. The Wolfenstein parametrization is
very useful for phenomenological applications, as we will see below. 
A detailed discussion of the next-to-leading order terms in $\lambda$ 
can be found in \cite{blo}.

\subsection{Unitarity Triangles}
The central targets for tests of the KM mechanism of CP violation are the
unitarity triangles of the CKM matrix. As we have already noted, the
CKM matrix is unitary. Consequently, it satisfies
\begin{equation}
\hat V_{\mbox{{\scriptsize CKM}}}^{\,\,\dagger}\cdot\hat 
V_{\mbox{{\scriptsize CKM}}}=
\hat 1=\hat V_{\mbox{{\scriptsize CKM}}}\cdot\hat V_{\mbox{{\scriptsize 
CKM}}}^{\,\,\dagger},
\end{equation}
implying a set of 12 equations, which consist of 6 normalization 
relations and 6 orthogonality relations. The latter can be represented as 
6 triangles in the complex plane \cite{AKL}, all having the same area, 
$2 A_{\Delta}=|J_{\rm CP}|$ \cite{UT-area}. However, in only two of them, 
all three sides are of comparable magnitude ${\cal O}(\lambda^3)$, while 
in the remaining ones, one side is suppressed with respect to the others by 
${\cal O}(\lambda^2)$ or ${\cal O}(\lambda^4)$. The orthogonality relations 
describing the non-squashed triangles are given by
\begin{eqnarray}
V_{ud}\,V_{ub}^\ast+V_{cd}\,V_{cb}^\ast+V_{td}\,V_{tb}^\ast&=&0
\quad\mbox{[1st and 3rd column]}\label{UT1}\\
V_{ub}^\ast\, V_{tb}+V_{us}^\ast\, V_{ts}+V_{ud}^\ast\, V_{td}&=&0
\quad\mbox{[1st and 3rd row]}.\label{UT2}
\end{eqnarray}
At leading order in $\lambda$, these relations agree with each other, and
yield
\begin{equation}\label{UTLO}
(\rho+i\eta)A\lambda^3+(-A\lambda^3)+(1-\rho-i\eta)A\lambda^3=0.
\end{equation}
Consequently, they describe the same triangle, which is usually referred to 
as {\it the} unitarity triangle of the CKM matrix \cite{UT-area,CK}. 
It is convenient to divide (\ref{UTLO}) by the overall normalization 
$A\lambda^3$. Then we obtain a triangle in the complex plane with a 
basis normalized to 1, and an apex given by $(\rho,\eta)$.

\begin{figure}[t]
\begin{tabular}{lr}
   \epsfysize=3.6cm
   \epsffile{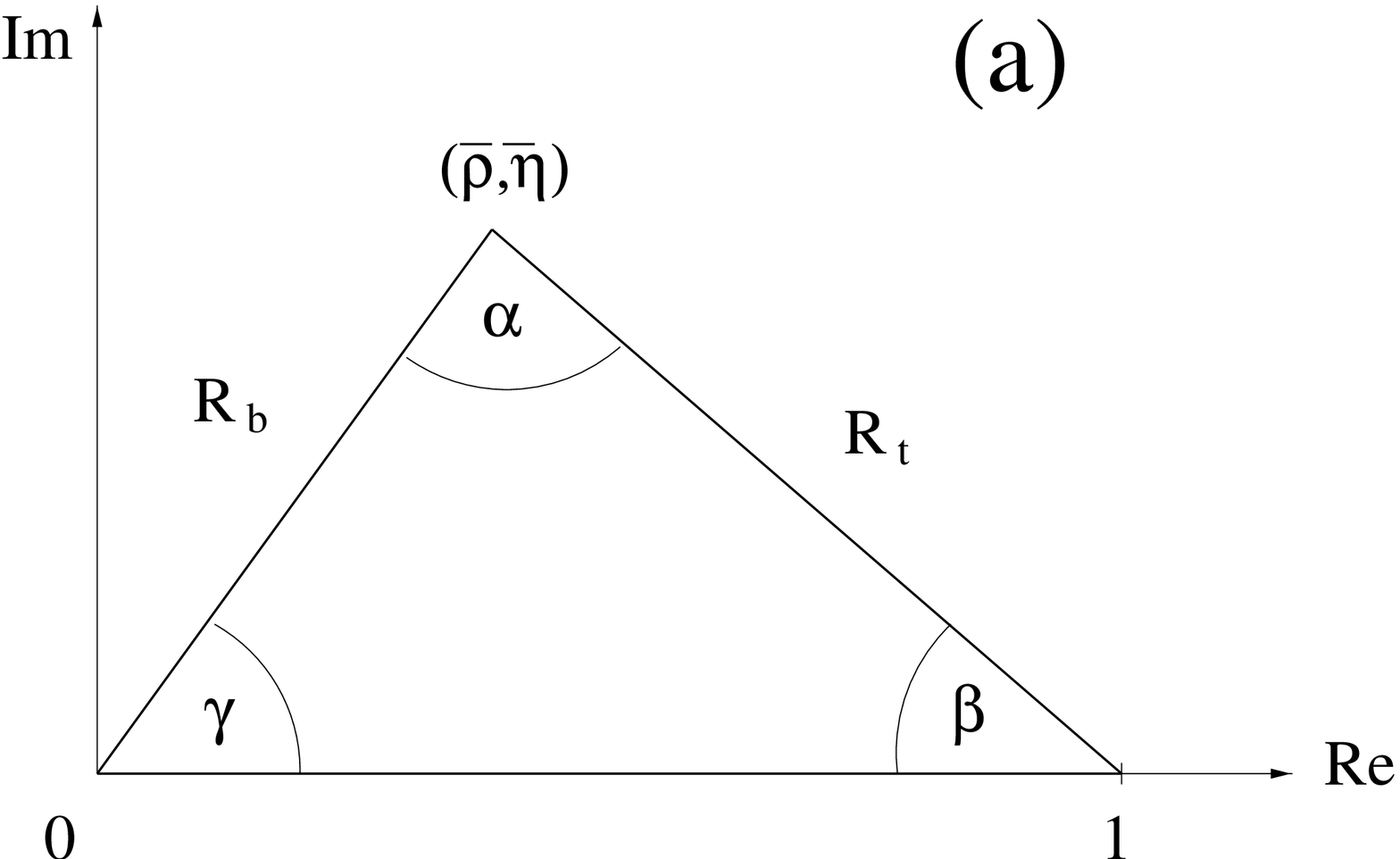}
&
   \epsfysize=3.6cm
   \epsffile{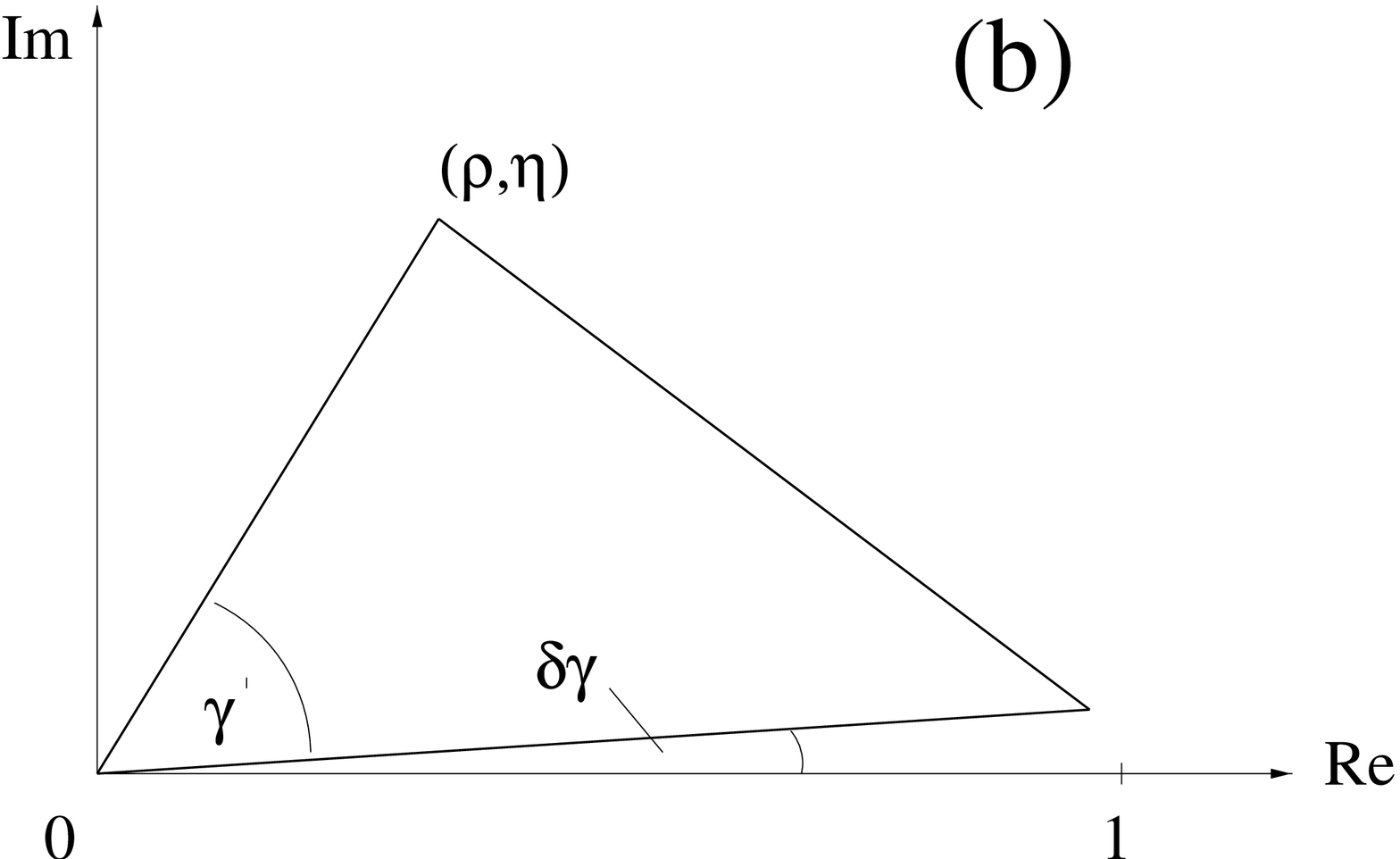}
\end{tabular}
\caption[]{The two non-squashed unitarity triangles of the CKM matrix: 
(a) and (b) correspond to the orthogonality relations (\ref{UT1}) and 
(\ref{UT2}), respectively.}
\label{fig:UT}
\end{figure}

In the future, the experimental accuracy will reach such an impressive 
level that we will have to distinguish between the unitarity triangles 
described by (\ref{UT1}) and (\ref{UT2}), which differ through 
${\cal O}(\lambda^2)$ corrections. They are illustrated in 
Fig.\ \ref{fig:UT}, where $\overline{\rho}$ and 
$\overline{\eta}$ are related to $\rho$ and $\eta$ through \cite{blo}
\begin{equation}
\overline{\rho}\equiv\left(1-\lambda^2/2\right)\rho,\quad
\overline{\eta}\equiv\left(1-\lambda^2/2\right)\eta,
\end{equation}
and 
\begin{equation}
\delta\gamma\equiv\gamma-\gamma'=\lambda^2\eta.
\end{equation}
The sides $R_b$ and $R_t$ of the unitarity triangle shown in 
Fig.\ \ref{fig:UT} (a) are given by
\begin{eqnarray}
R_b&=&\left(1-\frac{\lambda^2}{2}\right)\frac{1}{\lambda}\left|
\frac{V_{ub}}{V_{cb}}\right|\,=\,\sqrt{\overline{\rho}^2+\overline{\eta}^2}
\,=\,0.38\pm0.08\label{Rb-intro}\label{Rb-def}\\
R_t&=&\frac{1}{\lambda}\left|\frac{V_{td}}{V_{cb}}\right|\,=\,
\sqrt{(1-\overline{\rho})^2+\overline{\eta}^2}\,=\,{\cal O}(1),\label{Rt-def}
\end{eqnarray}
and will show up at several places throughout these lectures.
Whenever we refer to a unitarity triangle, we mean the one illustrated
in Fig.\ \ref{fig:UT} (a).

\begin{figure}
\vspace{0.10in}
\centerline{
\epsfysize=5.2truecm
\epsffile{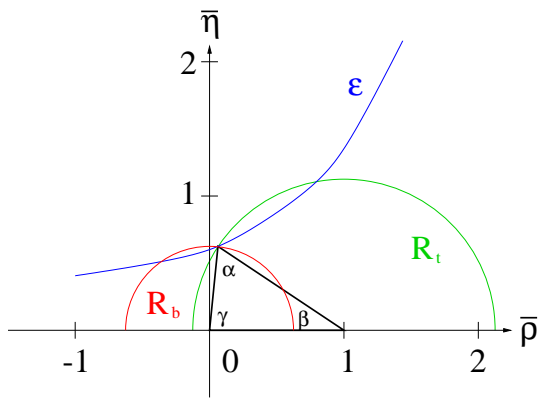}
}
\caption[]{Contours to determine the unitarity triangle in the
$\overline{\rho}$--$\overline{\eta}$ plane.}\label{fig:cont-scheme}
\end{figure}

\subsection{Standard Analysis of the Unitarity 
Triangle}\label{subsec:UT-analysis}
There is a ``standard analysis'' to constrain the apex of the unitarity
triangle in the $\overline{\rho}$--$\overline{\eta}$ plane, employing 
the following ingredients:
\begin{itemize}
\item Using heavy-quark arguments, exclusive and inclusive 
$b\to u,c\ell\overline{\nu}_\ell$ decays provide $|V_{ub}|$ and $|V_{cb}|$
\cite{ligeti}, allowing us to fix the side $R_b$ of the unitarity triangle, 
i.e.\ a circle in the $\overline{\rho}$--$\overline{\eta}$ plane around 
$(0,0)$ with radius $R_b$.
\item Using the top-quark mass $m_t$ as an input, and taking into account 
certain QCD corrections and non-perturbative parameters, we may extract
$|V_{td}|$ from $B^0_d$--$\overline{B^0_d}$ mixing (see below). The 
combination of $|V_{td}|$ with $|V_{cb}|$ allows us then to fix the side 
$R_t$ of the unitarity triangle, i.e.\ a 
circle in the $\overline{\rho}$--$\overline{\eta}$ plane around $(1,0)$ 
with radius $R_t$. Comparing $B^0_d$--$\overline{B^0_d}$ with
$B^0_s$--$\overline{B^0_s}$ mixing, an $SU(3)$-breaking parameter $\xi$
suffices to determine $R_t$.
\item Using $m_t$ and $|V_{cb}|$ as an input, and taking into account 
certain QCD corrections and non-perturbative parameters, the observable
$\varepsilon$ describing indirect CP violation in the kaon system allows
us to fix a hyperbola in the $\overline{\rho}$--$\overline{\eta}$ plane.
\end{itemize}
These contours are sketched in Fig.~\ref{fig:cont-scheme}; their 
intersection gives the apex of the unitarity triangle shown in 
Fig.\ \ref{fig:UT} (a). Because of strong correlations between theoretical 
and experimental uncertainties, it is rather involved to convert the
experimental information into an allowed range in the 
$\overline{\rho}$--$\overline{\eta}$ plane, and various analyses can
be found in the literature: a simple scanning approach \cite{buras-KAON},
a Gaussian approach \cite{al}, the ``BaBar 95\% scanning method'' 
\cite{Babar-95-scan}, a Bayesian approach \cite{bayes}, and a 
non-Bayesian statistical approach \cite{hoeck}. Other recent analyses 
can be found in \cite{AS-CKM,Bu-CKM}. A reasonable range for $\alpha$, 
$\beta$ and $\gamma$ that is consistent with these approaches is given by 
\begin{equation}\label{SM-ranges}
70^\circ\lsim\alpha\lsim130^\circ, \quad 20^\circ\lsim\beta\lsim30^\circ, 
\quad 50^\circ\lsim\gamma\lsim70^\circ.
\end{equation}
The question of how to combine the theoretical and experimental errors 
in an optimal way will certainly continue to be a hot topic in the future. 
This is also reflected by the Bayesian \cite{bayes} vs.\ 
non-Bayesian \cite{hoeck} debate going on at present.

\subsection{Quantitative Studies of CP Violation}
As we have seen above, the neutral kaon system provides two different 
CP-violating parameters, $\varepsilon$ and $\mbox{Re}(\varepsilon'/
\varepsilon)$. The former is one of the ingredients of the ``standard 
analysis'' of the unitarity triangle, implying in particular 
$\overline{\eta}>0$ if very plausible assumptions about a certain 
non-perturbative ``bag'' parameter are made. On the other hand, 
$\mbox{Re}(\varepsilon'/\varepsilon)$ does not (yet) provide further 
stringent constraints on the unitarity triangle because of large hadronic 
uncertainties, although the experimental values are of the same order 
of magnitude as the range of theoretical estimates 
\cite{bertolini,buras-KAON}. 

Considerably more promising in view of testing the Standard-Model
description of CP violation are the rare kaon decays 
$K^+\to\pi^+\nu\overline{\nu}$ and $K_{\rm L}\to\pi^0\nu\overline{\nu}$,
which originate in the Standard Model from loop effects and are 
theoretically very clean since the relevant hadronic matrix elements 
can be fixed through semileptonic kaon decays 
\cite{buras-KAON,isidori}. In particular, they also
allow an interesting determination of the unitarity triangle 
\cite{Kpinunu}, and show interesting correlations with CP violation
in the $B$ sector \cite{RF-Phys-Rep,BF-Kpinunu}. Unfortunately, the 
$K\to\pi\nu\overline{\nu}$ branching ratios are at the $10^{-11}$ level 
in the Standard Model; two events of $K^+\to\pi^+\nu\overline{\nu}$ have 
already been observed by the E787  Experiment at Brookhaven, yielding a 
branching ratio of $(1.57^{+1.75}_{-0.82})\times10^{-10}$ \cite{E787}. 
It is very important to measure $K^+\to\pi^+\nu\overline{\nu}$ and 
$K_{\rm L}\to\pi^0\nu\overline{\nu}$ with reasonable statistics, and 
there are efforts under way to accomplish this challenging 
goal \cite{Kpinunu-exp}.

In the case of the $B$-meson system, consisting of charged mesons 
$B^+_u\sim u\overline{b}$, $B^+_c\sim c\overline{b}$, as well as 
neutral ones $B^0_d\sim d\overline{b}$, $B^0_s\sim s\overline{b}$,
we have a ``simplified'' hadron dynamics, since the $b$ quark is 
``heavy'' with respect to the QCD scale parameter $\Lambda_{\rm QCD}$. 
Moreover, hadronic uncertainties can be eliminated or cancel in appropriate 
CP-violating observables, thereby providing various tests of the 
KM mechanism of CP violation and direct determinations of the angles 
of the unitarity triangle. As we will see below, the Standard Model 
predicts large CP-violating asymmetries in certain decays, and large 
effects were actually observed recently in $B_d\to J/\psi K_{\rm S}$ 
\cite{BaBar-CP-obs,Belle-CP-obs}. The goal is now to overconstrain the 
unitarity triangle as much as possible and to test several Standard-Model 
predictions, with the hope to encounter discrepancies that could shed 
light on the physics lying beyond the Standard Model. In this decade, 
the asymmetric $e^+e^-$ $B$ factories operating at the $\Upsilon(4S)$ 
resonance with their detectors BaBar and Belle provide access to several 
benchmark decay modes of $B^\pm_u$ and $B^0_d$ mesons \cite{BABAR-BOOK}. 
Moreover, experiments at hadron colliders allow us to study, in addition, 
large data samples of decays of $B_s$ mesons, which are another very 
important element in the testing of the Standard-Model description of 
CP violation. Important first steps in this direction are already expected 
at run II of the Tevatron \cite{TEV-BOOK}, whereas several strategies 
can only be fully exploited in the LHC era \cite{LHC-BOOK}, in particular 
at LHCb (CERN) and BTeV (Fermilab).

In these lectures we shall focus on the $B$-meson system. For the 
exploration of CP violation, non-leptonic $B$ decays play the central 
r\^ole, as CP-violating effects are due to certain interference effects 
that may show up in this decay class. Before turning to these modes, let 
us note that there are also other promising systems to obtain insights 
into CP violation, for example $D$ mesons, where the Standard Model 
predicts very small CP violation, electric dipole moments or hyperon 
decays. These topics are, however, beyond the scope of this presentation.

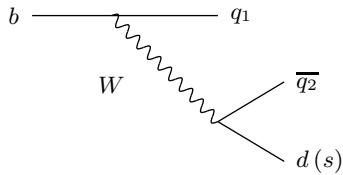
\begin{figure}[t]
\vspace*{-0.7truecm}
\begin{center}
{\small
\hspace*{4truecm}\begin{picture}(80,50)(80,20)
\Line(10,45)(80,45)\Photon(40,45)(80,5){2}{10}
\Line(80,5)(105,20)\Line(80,5)(105,-10)
\Text(5,45)[r]{$b$}\Text(85,45)[l]{$q_1$}
\Text(110,20)[l]{$\overline{q_2}$}
\Text(110,-10)[l]{$d\,(s)$}
\Text(45,22)[tr]{$W$}
\end{picture}}
\end{center}
\vspace*{1.0truecm}
\caption[]{Tree diagrams ($q_1,q_2\in\{u,c\}$).}\label{fig:tree-top}
\end{figure}

\begin{figure}[b]
\begin{center}
{\small
\begin{picture}(140,60)(0,20)
\Line(10,50)(130,50)\Text(5,50)[r]{$b$}\Text(140,50)[l]{$d\,(s)$}
\PhotonArc(70,50)(30,0,180){3}{15}
\Text(69,56)[b]{$u,c,t$}\Text(109,75)[b]{$W$}
\Gluon(70,50)(120,10){2}{10}
\Line(120,10)(135,23)\Line(120,10)(135,-3)
\Text(85,22)[tr]{$G$}\Text(140,-3)[l]{$q_1$}
\Text(140,23)[l]{$\overline{q_2}=\overline{q_1}$}
\end{picture}}
\end{center}
\vspace*{0.7truecm}
\caption[]{QCD penguin diagrams ($q_1=q_2\in\{u,d,c,s\}$).}\label{fig:QCD-top}
\end{figure}
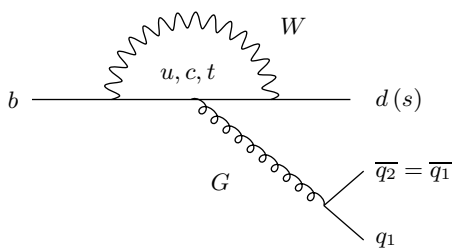

\section{Non-Leptonic $B$ Decays}\label{sec:non-lept}
\subsection{Classification}\label{sec:class}
Non-leptonic $\overline{B}$ decays are mediated by 
$b\to q_1\,\overline{q_2}\,d\,(s)$ quark-level transitions, with
$q_1,q_2\in\{u,d,c,s\}$. There are two kinds of topologies contributing 
to non-leptonic $B$ decays: tree-diagram-like and ``penguin'' topologies. 
The latter consist of gluonic (QCD) and electroweak (EW) penguins. 
In Figs.~\ref{fig:tree-top}--\ref{fig:EWP-top}, the corresponding 
leading-order Feynman diagrams are shown. Depending
on the flavour content of their final states, we may classify 
$b\to q_1\,\overline{q_2}\,d\,(s)$ decays as follows:
\begin{itemize}
\item $q_1\not=q_2\in\{u,c\}$: only tree diagrams contribute.
\item $q_1=q_2\in\{u,c\}$: tree and penguin diagrams contribute.
\item $q_1=q_2\in\{d,s\}$: only penguin diagrams contribute.
\end{itemize}

\subsection{Low-Energy Effective Hamiltonians}\label{subsec:ham}
In order to analyse non-leptonic $B$ decays theoretically, one uses 
low-energy effective Hamiltonians, which are calculated by making use 
of the operator product expansion, yielding transition matrix elements 
of the following structure:
\begin{equation}\label{ee2}
\langle f|{\cal H}_{\rm eff}|i\rangle=\frac{G_{\rm F}}{\sqrt{2}}
\lambda_{\rm CKM}\sum_k C_k(\mu)\langle f|Q_k(\mu)|i\rangle\,.
\end{equation}
The operator product expansion allows us to separate the short-distance
contributions to this transition amplitude from the long-distance 
ones, which are described by perturbative Wilson coefficient 
functions $C_k(\mu)$ and non-perturbative hadronic matrix elements 
$\langle f|Q_k(\mu)|i\rangle$, respectively. As usual, $G_{\rm F}$ is
the Fermi constant, $\lambda_{\rm CKM}$ is a CKM factor, and $\mu$ denotes an 
appropriate renormalization scale. The $Q_k$ are local operators, which 
are generated by electroweak interactions and QCD, and govern ``effectively'' 
the decay in question. The Wilson coefficients $C_k(\mu)$ can be 
considered as scale-dependent couplings related to the vertices described
by the $Q_k$.

\begin{figure}[t]
\vspace*{0.2truecm}
{\small
\begin{picture}(140,60)(0,20)
\Line(10,50)(130,50)\Text(5,50)[r]{$b$}\Text(140,50)[l]{$d\,(s)$}
\PhotonArc(70,50)(30,0,180){3}{15}
\Text(69,56)[b]{$u,c,t$}\Text(109,75)[b]{$W$}
\Gluon(70,50)(120,10){2}{10}
\Line(120,10)(135,23)\Line(120,10)(135,-3)
\Text(90,28)[tr]{$Z,\gamma$}\Text(140,-3)[l]{$q_1$}
\Text(140,23)[l]{$\overline{q_2}=\overline{q_1}$}
\end{picture}}
\hspace*{1.2truecm}
{\small
\begin{picture}(140,60)(0,20)
\Line(10,65)(130,65)\Text(5,65)[r]{$b$}\Text(140,65)[l]{$d\,(s)$}
\PhotonArc(70,65)(20,180,360){3}{10}
\Text(69,71)[b]{$u,c,t$}\Text(45,35)[b]{$W$}
\Photon(73,47)(120,10){2}{10}
\Line(120,10)(135,23)\Line(120,10)(135,-3)
\Text(95,25)[tr]{$Z,\gamma$}\Text(140,-3)[l]{$q_1$}
\Text(140,23)[l]{$\overline{q_2}=\overline{q_1}$}
\end{picture}}
\vspace*{1.0truecm}
\caption[]{Electroweak penguin diagrams 
($q_1=q_2\in\{u,d,c,s\}$).}\label{fig:EWP-top}
\end{figure}
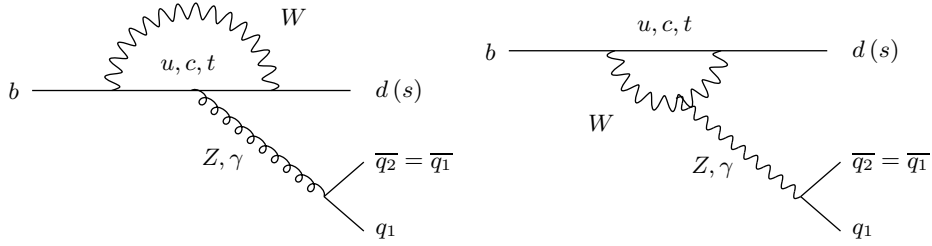

Let us consider $\overline{B^0_d}\to D^+K^-$, which is a pure ``tree'' 
decay, to discuss the evaluation of the corresponding low-energy effective
Hamiltonian in more detail. At leading order, this decay originates 
from a $b\to c \overline{u}s$ quark-level transition, where the $bc$ 
and $\overline{u}s$ quark currents are connected through the exchange of 
a $W$ boson. Evaluating the corresponding Feynman diagram yields 
\begin{equation}\label{trans-ampl}
-\,\frac{g_2^2}{8}V_{us}^\ast V_{cb}
\left[\overline{s}\gamma^\nu(1-\gamma_5)u\right]
\left[\frac{g_{\nu\mu}}{k^2-M_W^2}\right]
\left[\overline{c}\gamma^\mu(1-\gamma_5)b\right].
\end{equation}
Because of $k^2\approx m_b^2\ll M_W^2$, we have 
\begin{equation}
\frac{g_{\nu\mu}}{k^2-M_W^2}\quad\longrightarrow\quad
-\,\frac{g_{\nu\mu}}{M_W^2}\equiv-\left(\frac{8G_{\rm F}}{\sqrt{2}g_2^2}
\right)g_{\nu\mu},
\end{equation}
i.e.\ we may ``integrate out'' the $W$ boson in (\ref{trans-ampl}), 
and arrive at
\begin{eqnarray}
\lefteqn{{\cal H}_{\rm eff}=\frac{G_{\rm F}}{\sqrt{2}}V_{us}^\ast V_{cb}
\left[\overline{s}_\alpha\gamma_\mu(1-\gamma_5)u_\alpha\right]
\left[\overline{c}_\beta\gamma^\mu(1-\gamma_5)b_\beta\right]}\nonumber\\
&&=\frac{G_{\rm F}}{\sqrt{2}}V_{us}^\ast V_{cb}
(\overline{s}_\alpha u_\alpha)_{\mbox{{\scriptsize 
V--A}}}(\overline{c}_\beta b_\beta)_{\mbox{{\scriptsize V--A}}}
\equiv\frac{G_{\rm F}}{\sqrt{2}}V_{us}^\ast V_{cb}O_2\,,
\end{eqnarray}
where $\alpha$ and $\beta$ denote $SU(3)_{\rm C}$ colour indices. 
Effectively, our decay process $b\to c \overline{u}s$ is now
described by the ``current--current'' operator $O_2$. 

If we take into account QCD corrections, operator mixing leads to a second
``current--current'' operator, which is given by
\begin{equation}
O_1\equiv\left[\overline{s}_\alpha\gamma_\mu(1-\gamma_5)u_\beta\right]
\left[\overline{c}_\beta\gamma^\mu(1-\gamma_5)b_\alpha\right].
\end{equation}
Consequently, we obtain a low-energy effective Hamiltonian of the following
structure:
\begin{equation}\label{Heff-example}
{\cal H}_{\rm eff}=\frac{G_{\rm F}}{\sqrt{2}}V_{us}^\ast V_{cb}
\left[C_1(\mu)O_1+C_2(\mu)O_2\right],
\end{equation}
where $C_1(\mu)\not=0$ and $C_2(\mu)\not=1$ are due to QCD renormalization
effects. In order to evaluate these coefficients, we have first to 
calculate QCD corrections to the decay processes both in the full theory, 
i.e. with $W$ exchange, and in the effective theory, and have then to express 
the QCD-corrected transition amplitude in terms of QCD-corrected matrix 
elements and Wilson coefficients as in (\ref{ee2}). This procedure is 
called ``matching''. The results for the $C_k(\mu)$ thus obtained contain 
terms of $\mbox{log}(\mu/M_W)$, which become large for $\mu={\cal O}(m_b)$, 
the scale governing the hadronic matrix elements of the $O_k$. Making use of 
the renormalization group, which exploits the fact that the transition 
amplitude (\ref{ee2}) cannot depend on the chosen renormalization scale 
$\mu$, we may sum up the following terms of the Wilson coefficients:
\begin{equation}
\alpha_s^n\left[\log\left(\frac{\mu}{M_W}\right)\right]^n 
\,\,\mbox{(LO)},\quad\,\,\alpha_s^n\left[\log\left(\frac{\mu}{M_W}\right)
\right]^{n-1}\,\,\mbox{(NLO)},\quad ...
\end{equation}
A very detailed discussion of these techniques can be found in 
\cite{BBL-rev}.

In the case of decays receiving contributions both from tree and
from penguin topologies, basically the only difference to 
(\ref{Heff-example}) is that we encounter more operators:
\begin{equation}\label{e4}
{\cal H}_{\mbox{{\scriptsize eff}}}=\frac{G_{\mbox{{\scriptsize 
F}}}}{\sqrt{2}}\left[\sum\limits_{j=u,c}V_{jr}^\ast V_{jb}\left\{\sum
\limits_{k=1}^2C_k(\mu)\,Q_k^{jr}+\sum\limits_{k=3}^{10}C_k(\mu)\,Q_k^{r}
\right\}\right].
\end{equation}
Here the current--current operators $Q_{1}^{jr}$ and $Q_{2}^{jr}$,
the QCD penguin operators $Q_{3}^r$--$Q_{6}^r$, and the EW penguin operators
$Q_{7}^r$--$Q_{10}^r$ are related to the tree, QCD and EW penguin processes 
shown in Figs.~\ref{fig:tree-top}--\ref{fig:EWP-top} (explicit expressions
for these operators can be found in \cite{RF-Phys-Rep,BBL-rev}).
At a renormalization scale $\mu={\cal O}(m_b)$, the Wilson coefficients of 
the current--current operators satisfy $C_1(\mu)={\cal O}(10^{-1})$ and 
$C_2(\mu)={\cal O}(1)$, whereas those of the penguin operators are 
${\cal O}(10^{-2})$. Note that penguin topologies with internal charm- 
and up-quark exchanges are described in this framework by penguin-like 
matrix elements of the corresponding current--current operators 
\cite{RF-DIPL}, and may also have important phenomenological consequences 
\cite{BF-PEN,CHARM-PEN}.

Since the ratio $\alpha/\alpha_s={\cal O}(10^{-2})$ of the QED and QCD 
couplings is very small, we would expect na\"\i vely that EW penguins 
should play a minor r\^ole in comparison with QCD penguins. This would 
actually be the case if the top quark was not ``heavy''. However, since 
the Wilson coefficient $C_9$ increases strongly with $m_t$, we obtain 
interesting EW penguin effects in several $B$ decays: $B^-\to K^-\phi$ 
is affected significantly by EW penguins, whereas $B\to\pi\phi$ and 
$B_s\to\pi^0\phi$ are even dominated by such topologies 
\cite{RF-EWP,RF-rev}. EW penguins also have an important impact on 
$B\to\pi K$ modes \cite{EWP-BpiK}, as we will see in Section~\ref{sec:recent}. 

The low-energy effective Hamiltonians discussed in this section apply to 
all $B$ decays that are caused by the same corresponding quark-level 
transition, i.e.\ they are ``universal''. Within this formalism, differences 
between various exclusive modes are only due to the hadronic matrix elements 
of the relevant four-quark operators. Unfortunately, the evaluation of such 
matrix elements is associated with large uncertainties and is a very 
challenging task. In this context, ``factorization'' is a widely used 
concept, which is our next topic.

\subsection{Factorization of Hadronic Matrix Elements}
In order to discuss ``factorization'', let us consider once more 
$\overline{B^0_d}\to D^+K^-$. Evaluating the corresponding 
transition amplitude, we encounter the hadronic matrix elements of the 
$O_{1,2}$ operators between the $\langle K^-D^+|$ final and 
$|\overline{B^0_d}\rangle$ initial states. If we use the well-known 
$SU(N_{\rm C})$ colour-algebra relation
\begin{equation}
T^a_{\alpha\beta}T^a_{\gamma\delta}=\frac{1}{2}\left(\delta_{\alpha\delta}
\delta_{\beta\gamma}-\frac{1}{N_{\rm C}}\delta_{\alpha\beta}
\delta_{\gamma\delta}\right)
\end{equation}
to rewrite the operator $O_1$, we obtain
\begin{displaymath}
\langle K^-D^+|{\cal H}_{\rm eff}|\overline{B^0_d}\rangle=
\frac{G_{\rm F}}{\sqrt{2}}V_{us}^\ast V_{cb}\Bigl[a_1\langle K^-D^+|
(\overline{s}_\alpha u_\alpha)_{\mbox{{\scriptsize V--A}}}
(\overline{c}_\beta b_\beta)_{\mbox{{\scriptsize V--A}}}
|\overline{B^0_d}\rangle
\end{displaymath}
\vspace*{-0.3truecm}
\begin{equation}
+2\,C_1\langle K^-D^+|
(\overline{s}_\alpha\, T^a_{\alpha\beta}\,u_\beta)_{\mbox{{\scriptsize 
V--A}}}(\overline{c}_\gamma 
\,T^a_{\gamma\delta}\,b_\delta)_{\mbox{{\scriptsize V--A}}}
|\overline{B^0_d}\rangle\Bigr],\nonumber
\end{equation}
with
\begin{equation}\label{a1-def}
a_1=\frac{C_1}{N_{\rm C}}+C_2.
\end{equation}
It is now straightforward to ``factorize'' the hadronic matrix elements:
\begin{eqnarray}
\lefteqn{\left.\langle K^-D^+|
(\overline{s}_\alpha u_\alpha)_{\mbox{{\scriptsize 
V--A}}}(\overline{c}_\beta b_\beta)_{\mbox{{\scriptsize V--A}}}
|\overline{B^0_d}\rangle\right|_{\rm fact}}\nonumber\\
&&=\langle K^-|\left[\overline{s}_\alpha\gamma_\mu(1-\gamma_5)u_\alpha\right]
|0\rangle\langle D^+|\left[\overline{c}_\beta\gamma^\mu
(1-\gamma_5)b_\beta\right]|\overline{B^0_d}\rangle\nonumber\\
&&\propto f_K \mbox{(``decay constant'')}\times 
F_{BD} \mbox{(``form factor'')},
\end{eqnarray}
\begin{equation}
\left.\langle K^-D^+|
(\overline{s}_\alpha\, T^a_{\alpha\beta}\,u_\beta)_{\mbox{{\scriptsize 
V--A}}}(\overline{c}_\gamma 
\,T^a_{\gamma\delta}\,b_\delta)_{\mbox{{\scriptsize V--A}}}
|\overline{B^0_d}\rangle\right|_{\rm fact}=0.
\end{equation}
The quantity introduced in (\ref{a1-def}) is a phenomenological
``colour factor'', governing ``colour-allowed'' decays. In the case of
``colour-suppressed'' modes, for instance $\overline{B^0_d}\to
\pi^0D^0$, we have to deal with the combination
\begin{equation}\label{a2-def}
a_2=C_1+\frac{C_2}{N_{\rm C}}.
\end{equation}

The concept of the factorization of hadronic matrix elements has
a long history \cite{Neu-Ste}, and can be justified, for example, 
in the large $N_{\rm C}$ limit \cite{largeN}. Recently, the
``QCD factorization'' approach was developed \cite{BBNS1,BBNS2,BBNS3}, which
may provide an important step towards a rigorous basis for factorization 
for a large class of non-leptonic two-body $B$-meson decays in the 
heavy-quark limit. The resulting formula for the transition amplitudes 
incorporates elements both of the na\"\i ve factorization approach 
sketched above and of the hard-scattering picture. Let us consider a decay 
$\overline{B}\to M_1M_2$, where $M_1$ picks up the spectator quark. 
If $M_1$ is either a heavy ($D$) or a light ($\pi$, $K$) meson, and 
$M_2$ a light ($\pi$, $K$) meson, QCD factorization gives a transition 
amplitude of the following structure:
\begin{equation}
A(\overline{B}\to M_1M_2)=\left[\mbox{``na\"\i ve factorization''}\right]
\times\left[1+{\cal O}(\alpha_s)+{\cal O}(\Lambda_{\rm QCD}/m_b)\right].
\end{equation}
While the ${\cal O}(\alpha_s)$ terms, i.e.\ the radiative 
non-factorizable corrections to na\"\i ve factorization, can be 
calculated in a systematic way, the main limitation of the 
theoretical accuracy is due to the ${\cal O}(\Lambda_{\rm QCD}/m_b)$ 
terms. These issues are discussed in detail in \cite{neubert}.
Further interesting recent papers are listed in \cite{fact-recent}.

Another QCD approach to deal with non-leptonic $B$ decays into charmless
final states -- the perturbative hard-scattering (or ``PQCD'') approach --
was developed independently in \cite{PQCD}, and differs from the QCD
factorization formalism in some technical aspects. An interesting 
avenue to deal with non-leptonic $B$ decays is also provided by
QCD light-cone sum-rule approaches \cite{sum-rules}.

\section{Towards Studies of CP Violation in the $B$ 
System}\label{sec:cp-strat}
\subsection{Amplitude Structure and Direct CP Violation}
If we use the unitarity of the CKM matrix, it is an easy exercise to
show that the amplitude for any given non-leptonic $B$ decay can always
be written is such a way that at most two weak CKM amplitudes contribute:
\begin{eqnarray}
A(\overline{B}\to\overline{f})&=&e^{+i\varphi_1}
|A_1|e^{i\delta_1}+e^{+i\varphi_2}|A_2|e^{i\delta_2}\label{par-ampl}\\
A(B\to f)&=&e^{-i\varphi_1}|A_1|e^{i\delta_1}+
e^{-i\varphi_2}|A_2|e^{i\delta_2}.\label{par-ampl-CP}
\end{eqnarray}
Here $\varphi_{1,2}$ denote CP-violating weak phases, which are due to
the CKM matrix, and the $|A_{1,2}|e^{i\delta_{1,2}}$ are CP-conserving 
``strong'' amplitudes, containing the whole hadron dynamics of the 
decay at hand:
\begin{equation}
|A|e^{i\delta}\sim\sum\limits_k
\underbrace{C_{k}(\mu)}_{\mbox{pert.\ QCD}} 
\times\,\,\, \underbrace{\langle\overline{f}|Q_{k}(\mu)
|\overline{B}\rangle}_{\mbox{non-pert.}}.
\end{equation}
Employing (\ref{par-ampl}) and (\ref{par-ampl-CP}), we obtain the following
CP-violating rate asymmetry:
\begin{eqnarray}
{\cal A}_{\rm CP}&\equiv&\frac{\Gamma(B\to f)-
\Gamma(\overline{B}\to\overline{f})}{\Gamma(B\to f)+\Gamma(\overline{B}
\to \overline{f})}=
\frac{|A(B\to f)|^2-|A(\overline{B}\to \overline{f})|^2}{|A(B\to f)|^2+
|A(\overline{B}\to \overline{f})|^2}\nonumber\\
&=&\frac{2|A_1||A_2|\sin(\delta_1-\delta_2)
\sin(\varphi_1-\varphi_2)}{|A_1|^2+2|A_1||A_2|\cos(\delta_1-\delta_2)
\cos(\varphi_1-\varphi_2)+|A_2|^2}.\label{direct-CPV}
\end{eqnarray}
Consequently, a non-vanishing CP asymmetry ${\cal A}_{\rm CP}$ arises from 
interference effects between the two weak amplitudes, and requires both
a non-trivial weak phase difference $\varphi_1-\varphi_2$ and a
non-trivial strong phase difference $\delta_1-\delta_2$. This kind of
CP violation is referred to as ``direct'' CP violation, as it originates
directly at the amplitude level of the considered decay. It is the
$B$-meson counterpart of the effects probed through 
$\mbox{Re}(\varepsilon'/\varepsilon)$ in the neutral kaon system.
Since $\varphi_1-\varphi_2$ is in general given by one of the angles of 
the unitarity triangle -- usually $\gamma$ -- the goal is to determine this 
quantity from the measured value of ${\cal A}_{\rm CP}$. Unfortunately, 
the extraction of $\varphi_1-\varphi_2$ from ${\cal A}_{\rm CP}$ is affected 
by hadronic uncertainties, which are due to the strong 
amplitudes $|A_{1,2}|e^{i\delta_{1,2}}$ (see (\ref{direct-CPV})).

\subsection{Classification of the Main Strategies}\label{sec:strat}
The most obvious -- but also most challenging -- strategy we may follow
is to try to calculate the relevant hadronic matrix elements
$\langle \overline{f}|Q_k(\mu)|\overline{B}\rangle$. As we have noted
above, interesting progress has recently been made in this direction
through the development of the QCD factorization 
\cite{BBNS1,BBNS2,BBNS3,fact-recent}, the PQCD \cite{PQCD}, and the QCD 
light-cone sum-rule approaches \cite{sum-rules}.

\begin{figure}[t]
\begin{center}
\begin{picture}(320,150)(0,0)
\Line(50,10)(208,10) \ArrowLine(208,10)(210,10)
\DashLine(50,10)(290,130){6}\ArrowLine(288,129)(290,130)
\DashLine(210,10)(290,130){6}\ArrowLine(289,129)(290,130)
\Line(50,10)(130,130)\ArrowLine(128.3,128)(129,129)
\Line(210,10)(130,130)\ArrowLine(131.5,128)(131,129)
\Text(130,2)[t]{$A(B^+_c\to D_s^+\overline{D^0})=A(B^-_c\to D^-_sD^0)$}
\Text(82,80)[r]{$\sqrt{2}\,A(B^+_c\to D_s^+ D_+^0)$}
\Text(255,60)[l]{$A(B^-_c\to D^-_s\overline{D^0})$}
\Text(310,140)[br]{$\sqrt{2}\,A(B^-_c\to D_s^- D_+^0)$}
\Line(252,138)(262,118)\ArrowLine(261,120)(262,118)
\Text(95,140)[lb]{$A(B^+_c\to D^+_sD^0)$}
\Line(150,138)(145,110)\ArrowLine(145.5,113)(145,110)
\Text(210,29)[c]{$2\gamma$}\CArc(210,10)(33,57,123)
\ArrowLine(145.5,113)(145,110)
\end{picture}
\end{center}
\caption{The extraction of $\gamma$ from $B_c^\pm\to 
D^\pm_s\{D^0,\overline{D^0},D^0_+\}$ decays.}\label{Bc-triangles}
\end{figure}
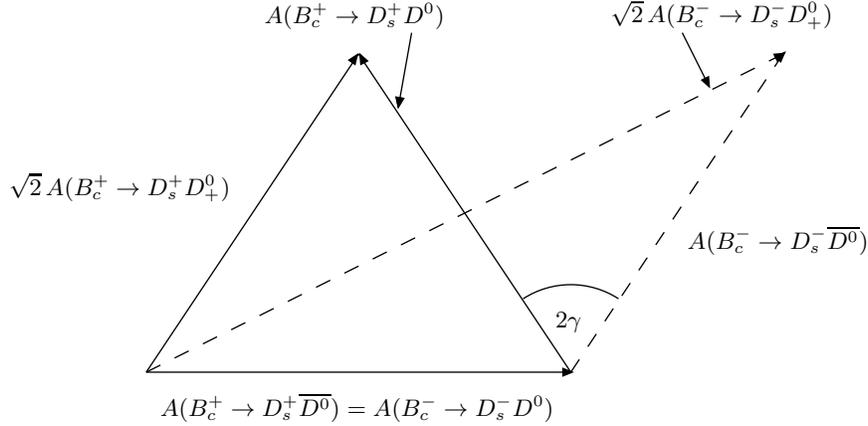

Another avenue we may follow is to search for fortunate cases, where
relations between decay amplitudes allow us to eliminate the hadronic
uncertainties. This approach was pioneered by Gronau and Wyler \cite{gw}, 
who proposed the extraction of $\gamma$ from triangle relations between
$B^\pm_u\to K^\pm \{D^0,\overline{D^0},D^0_+\}$ amplitudes, where
$D^0_+$ is the CP-even eigenstate of the neutral $D$-meson system.
These modes receive only contributions from tree-diagram-like topologies. 
Unfortunately, this strategy, which is {\it theoretically clean}, is very 
difficult from an experimental point 
of view, since the corresponding triangles are very squashed ones
(for other experimental problems and strategies to solve them, see
\cite{ADS}). As an alternative $B_d\to K^{\ast0}\{D^0,\overline{D^0},D^0_+\}$ 
modes were proposed \cite{dun}, where the triangles are more equilateral. 
Interestingly, from a theoretical point of view, the ideal realization of 
this ``triangle'' approach arises in the $B_c$-meson system. Here the 
$B^\pm_c\to D_s^\pm \{D^0,\overline{D^0},D^0_+\}$ decays allow
us to construct the amplitude triangles sketched in Fig.~\ref{Bc-triangles}, 
where all sides are expected to be of the same order of magnitude \cite{FW}.
The practical implementation of this strategy appears also to be challenging, 
but elaborate feasibility studies for experiments of the LHC era are strongly 
encouraged. Amplitude relations can also be derived with the help of the
flavour symmetries of strong interactions, i.e.\ $SU(2)$ and $SU(3)$. 
Here we have to deal with $B_{(s)}\to \pi\pi, \pi K, KK$ decays, providing
interesting determinations of weak phases and insights into hadronic physics. 
We shall have a closer look at these modes in Section~\ref{sec:recent}.

\begin{figure}[t]
\begin{center}
{\small
\begin{picture}(250,70)(0,45)
\ArrowLine(10,100)(40,100)\Photon(40,100)(80,100){2}{8}
\ArrowLine(80,100)(110,100)
\ArrowLine(40,60)(10,60)\Photon(40,60)(80,60){2}{8}
\ArrowLine(110,60)(80,60)
\ArrowLine(40,100)(40,60)\ArrowLine(80,60)(80,100)
\Text(25,105)[b]{$q$}\Text(60,105)[b]{$W$}\Text(95,105)[b]{$b$}
\Text(25,55)[t]{$b$}\Text(60,55)[t]{$W$}\Text(95,55)[t]{$q$}
\Text(35,80)[r]{$u,c,t$}\Text(85,80)[l]{$u,c,t$}
\ArrowLine(160,100)(190,100)\ArrowLine(190,100)(230,100)
\ArrowLine(230,100)(260,100)
\ArrowLine(190,60)(160,60)\ArrowLine(230,60)(190,60)
\ArrowLine(260,60)(230,60)
\Photon(190,100)(190,60){2}{8}\Photon(230,60)(230,100){2}{8}
\Text(175,105)[b]{$q$}\Text(245,105)[b]{$b$}
\Text(175,55)[t]{$b$}\Text(245,55)[t]{$q$}
\Text(210,105)[b]{$u,c,t$}\Text(210,55)[t]{$u,c,t$}
\Text(180,80)[r]{$W$}\Text(240,80)[l]{$W$}
\end{picture}}
\end{center}
\vspace*{-0.4truecm}
\caption{Box diagrams contributing to $B^0_q$--$\overline{B^0_q}$ mixing 
($q\in\{d,s\}$).}\label{fig:boxes}
\end{figure}
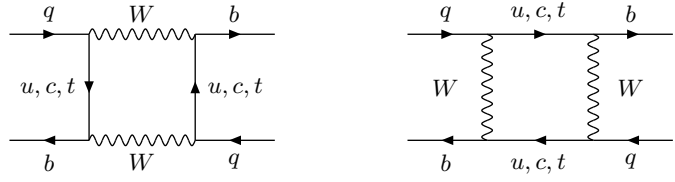

The third avenue we may follow to deal with the problems arising from
hadronic matrix elements is to employ decays of neutral $B_d$ or
$B_s$ mesons. Here we encounter a new kind of CP violation, which is
due to interference effects between $B^0_q$--$\overline{B^0_q}$ mixing 
and decay processes, and is referred to as ``mixing-induced'' 
CP violation. Within the Standard Model, $B^0_q$--$\overline{B^0_q}$ mixing
arises from the box diagrams shown in Fig.~\ref{fig:boxes}. Because of
this phenomenon, an initially, i.e.\ at time $t=0$, present 
$B^0_q$-meson state evolves into a time-dependent linear combination of 
$B^0_q$ and $\overline{B^0_q}$ states:
\begin{equation}
|B_q(t)\rangle=a(t)|B^0_q\rangle + b(t)|\overline{B^0_q}\rangle,
\end{equation}
where $a(t)$ and $b(t)$ are governed by an appropriate Schr\"odinger
equation. In order to solve it, mass eigenstates with mass and
width differences 
\begin{equation}
\Delta M_q\equiv M_{\rm H}^{(q)}-M_{\rm L}^{(q)}>0 \quad\mbox{and}\quad
\Delta\Gamma_q\equiv\Gamma_{\rm H}^{(q)}
-\Gamma_{\rm L}^{(q)},
\end{equation}
respectively, are introduced. The decay rates 
$\Gamma(\stackrel{{\mbox{\tiny (---)}}}{B^0_q}(t)\to 
\stackrel{{\mbox{\tiny (---)}}}{f})$ then contain terms proportional
to $\cos(\Delta M_qt)$ and $\sin(\Delta M_qt)$, describing the
$B^0_q$--$\overline{B^0_q}$ oscillations. To be specific, let us 
consider the very important special case where the $B_q^0$ meson 
decays into a final CP eigenstate $f$, satisfying
\begin{equation}\label{CP-rel}
(CP)|f\rangle=\pm |f\rangle.
\end{equation}
The corresponding time-dependent CP asymmetry then takes the following 
form:
\begin{eqnarray}
\lefteqn{a_{\rm CP}(t)\equiv\frac{\Gamma(B^0_q(t)\to f)-
\Gamma(\overline{B^0_q}(t)\to f)}{\Gamma(B^0_q(t)\to f)+
\Gamma(\overline{B^0_q}(t)\to f)}}\label{ee6}\\
&&=\left[\frac{{\cal A}_{\rm CP}^{\rm dir}(B_q\to f)\,\cos(\Delta M_q t)+
{\cal A}_{\rm CP}^{\rm mix}(B_q\to f)\,\sin(\Delta 
M_q t)}{\cosh(\Delta\Gamma_qt/2)-{\cal A}_{\rm 
\Delta\Gamma}(B_q\to f)\,\sinh(\Delta\Gamma_qt/2)}\right].\nonumber
\end{eqnarray}
In order to calculate the CP-violating observables, it is convenient to 
introduce 
\begin{equation}
\xi_f^{(q)}=\pm e^{-i\Theta_{{\rm \,M}}^{(q)}}
\frac{A(\overline{B_q^0}\to \overline{f})}{A(B_q^0\to f)},
\end{equation}
where $\pm$ refers to the CP eigenvalue of the final state $f$ specified
in (\ref{CP-rel}), and 
\begin{equation}
\Theta_{{\rm \,M}}^{(q)}-\pi=
2\,\mbox{arg}(V_{tq}^\ast V_{tb})
\equiv \phi_q=\left\{\begin{array}{cl}
+2\beta={\cal O}(50^\circ) & \mbox{for $q=d$,}\\
-2\delta\gamma={\cal O}(-2^\circ) & 
\mbox{for $q=s$}
\end{array}\right.
\end{equation}
is the CP-violating weak $B^0_q$--$\overline{B^0_q}$ mixing phase.
It should be noted that $\xi_f^{(q)}$ does not depend on the chosen CP or
CKM phase conventions and is actually a physical observable (for a 
detailed discussion, see \cite{RF-Phys-Rep}). We then obtain 
\begin{equation}
{\cal A}^{\rm dir}_{\rm CP}(B_q\to f)=
\frac{1-\bigl|\xi_f^{(q)}\bigr|^2}{1+\bigl|\xi_f^{(q)}\bigr|^2}=
\frac{|A(B\to f)|^2-|A(\overline{B}\to \overline{f})|^2}{|A(B\to f)|^2+
|A(\overline{B}\to \overline{f})|^2},
\end{equation}
and conclude that this observable measures direct CP violation, which we 
have already encountered in (\ref{direct-CPV}). The interesting new aspect 
is ``mixing-induced'' CP violation, which is described by
\begin{equation}
{\cal A}^{\rm mix}_{\rm CP}(B_q\to f)=
\frac{2 \mbox{Im} \xi^{(q)}_f}{1+\bigl|\xi^{(q)}_f\bigr|^2},
\end{equation}
and arises from interference effects between $B_q^0$--$\overline{B_q^0}$ 
mixing and decay processes. The width difference $\Delta\Gamma_q$, which 
may be sizeable in the $B_s$ system, as we will see in 
Subsection~\ref{subsec:Bs-features}, provides another observable, 
\begin{equation}\label{ADGam}
{\cal A}_{\rm \Delta\Gamma}(B_q\to f)\equiv
\frac{2\,\mbox{Re}\,\xi^{(q)}_f}{1+\bigl|\xi^{(q)}_f
\bigr|^2},
\end{equation}
which is, however, not independent from 
${\cal A}^{\rm dir}_{\rm CP}(B_q\to f)$ and 
${\cal A}^{\rm mix}_{\rm CP}(B_q\to f)$:
\begin{equation}\label{Obs-rel}
\Bigl[{\cal A}_{\rm CP}^{\rm dir}(B_q\to f)\Bigr]^2+
\Bigl[{\cal A}_{\rm CP}^{\rm mix}(B_q\to f)\Bigr]^2+
\Bigl[{\cal A}_{\Delta\Gamma}(B_q\to f)\Bigr]^2=1.
\end{equation}
Let us now have a closer look at $\xi_f^{(q)}$. Using (\ref{par-ampl}) 
and (\ref{par-ampl-CP}), we obtain
\begin{equation}
\xi_f^{(q)}=\mp e^{-i\phi_q}\left[
\frac{e^{+i\varphi_1}|A_1|e^{i\delta_1}+
e^{+i\varphi_2}|A_2|e^{i\delta_2}}{e^{-i\varphi_1}|A_1|e^{i\delta_1}+
e^{-i\varphi_2}|A_2|e^{i\delta_2}}\right],
\end{equation}
and observe that the calculation of $\xi_f^{(q)}$ is in general affected
by hadronic uncertainties. However, if one CKM amplitude plays the 
dominant r\^ole, the corresponding hadronic matrix element cancels:
\begin{equation}
\xi_f^{(q)}=\mp e^{-i\phi_q}\left[\frac{e^{+i\phi_f/2}|M_f|
e^{i\delta_f}}{e^{-i\phi_f/2}|M_f|e^{i\delta_f}}\right]=
\mp e^{-i(\phi_q-\phi_f)}.
\end{equation}
In this special case, direct CP violation vanishes, i.e.\ 
${\cal A}^{\rm dir}_{\rm CP}(B_q\to f)=0$. However, we still have
mixing-induced CP violation, measuring the CP-violating weak phase
difference $\phi\equiv\phi_q-\phi_f$ {\it without} hadronic uncertainties:
\begin{equation}
{\cal A}^{\rm mix}_{\rm CP}(B_q\to f)=\pm\sin\phi.
\end{equation}
The corresponding time-dependent CP asymmetry now takes the following
simple form:
\begin{equation}
\left.\frac{\Gamma(B^0_q(t)\to f)-
\Gamma(\overline{B^0_q}(t)\to \overline{f})}{\Gamma(B^0_q(t)\to f)+
\Gamma(\overline{B^0_q}(t)\to \overline{f})}\right|_{\Delta\Gamma_q=0}
=\pm\sin\phi\,\sin(\Delta M_q t),
\end{equation}
and allows an elegant determination of $\sin\phi$. Let us apply this
formalism, in the next section, to important benchmark modes for the
$B$ factories.

\begin{figure}[t]
\begin{center}
\leavevmode
\epsfysize=4.0truecm 
\epsffile{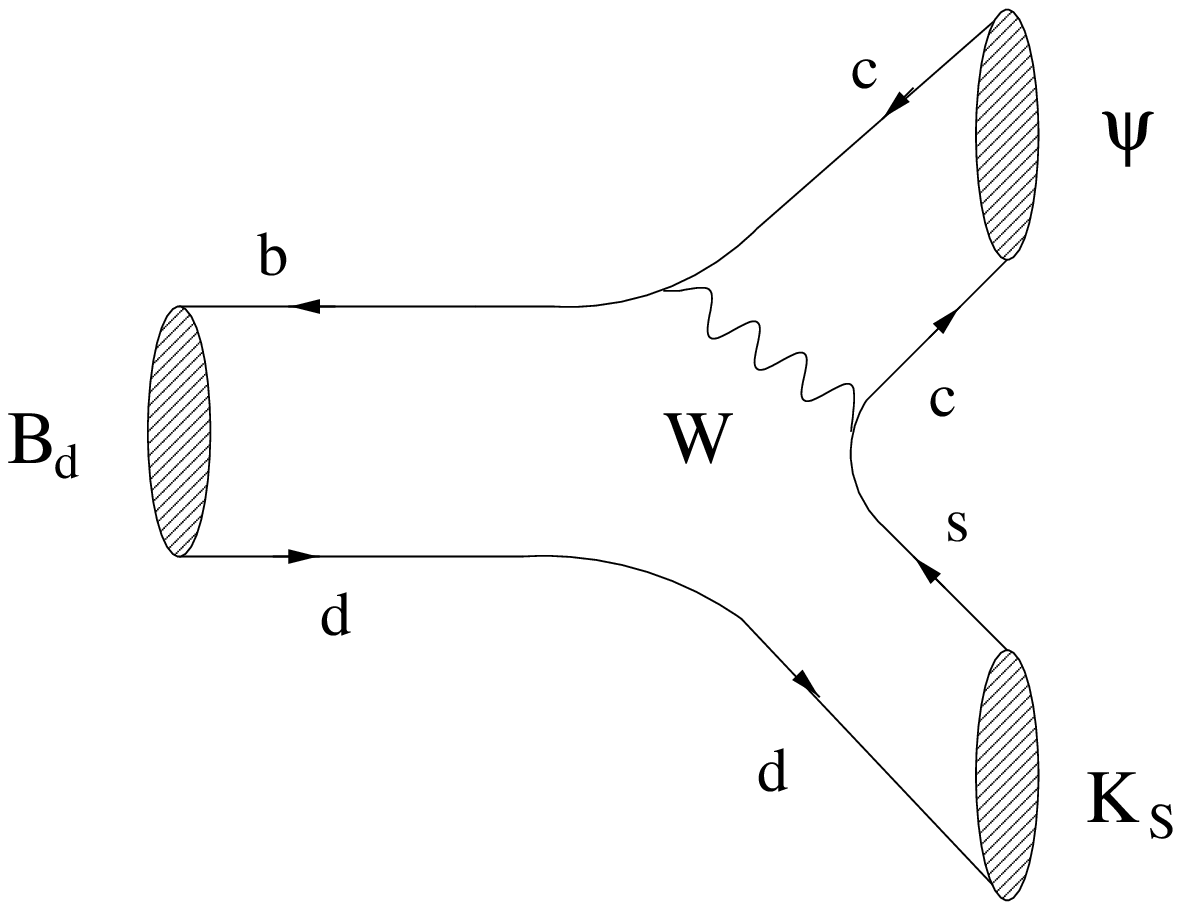} \hspace*{1truecm}
\epsfysize=4.0truecm 
\epsffile{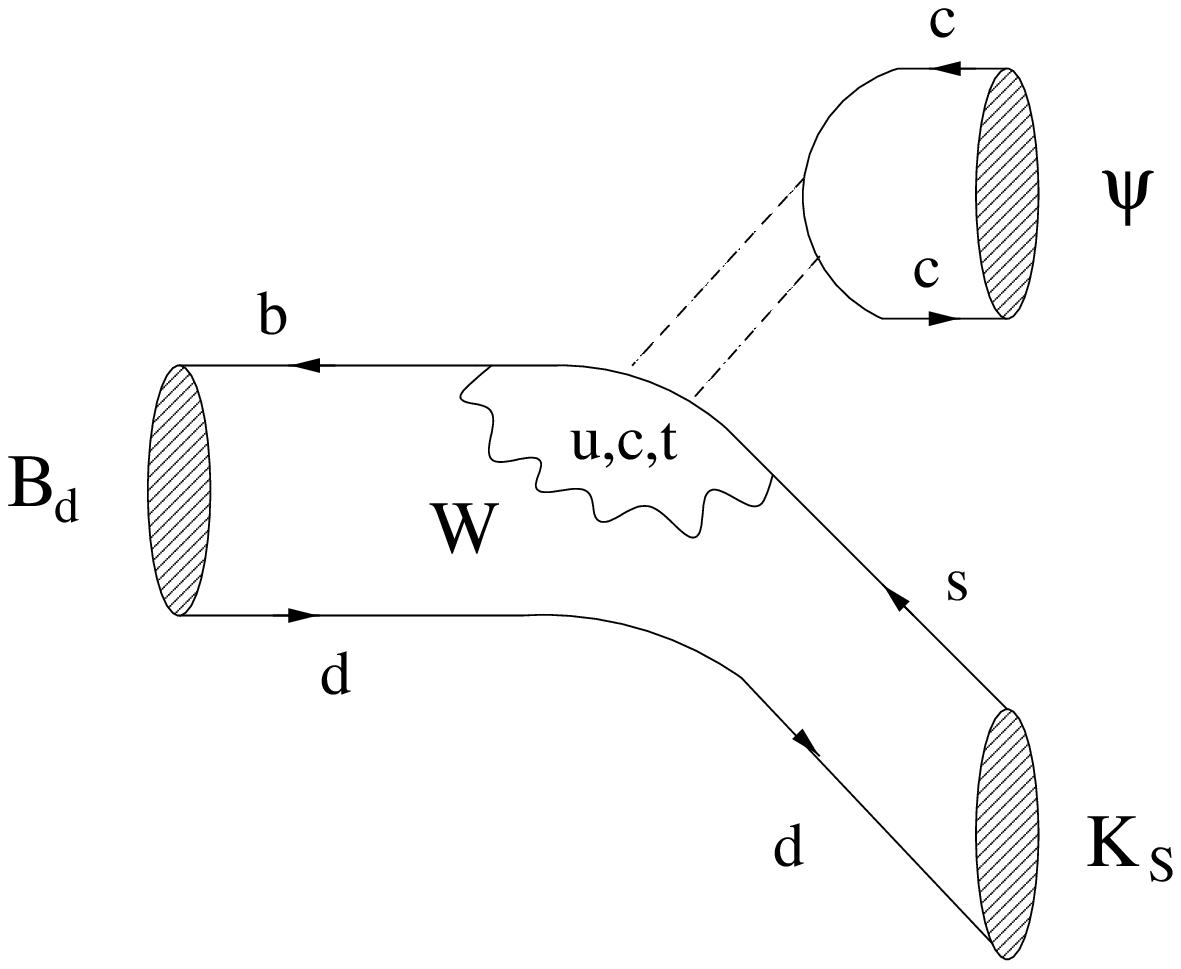}
\end{center}
\vspace*{-0.3truecm}
\caption{Feynman diagrams contributing to $B_d^0\to J/\psi K_{\rm S}$.
The dashed lines in the penguin topology represent a colour-singlet 
exchange.}\label{fig:BdPsiKS}
\end{figure}

\section{Benchmark Modes for the $B$ Factories}\label{sec:b-fact-bench}
\subsection{The ``Gold-Plated'' Mode $B_d\to J/\psi K_{\rm S}$}
The decay $B_d^0\to J/\psi\,K_{\rm S}$ is a transition into a CP 
eigenstate with eigenvalue $-1$, and originates from $\overline{b}\to
\overline{c} c \overline{s}$ quark-level decays. As can be seen in 
Fig.~\ref{fig:BdPsiKS}, we have to 
deal both with tree-diagram-like and with penguin topologies. The 
corresponding amplitude can be written as \cite{RF-BdsPsiK}
\begin{equation}\label{Bd-ampl1}
A(B_d^0\to J/\psi K_{\rm S})=\lambda_c^{(s)}\left(A_{\rm CC}^{c'}+
A_{\rm pen}^{c'}\right)+\lambda_u^{(s)}A_{\rm pen}^{u'}
+\lambda_t^{(s)}A_{\rm pen}^{t'}\,,
\end{equation}
where $A_{\rm CC}^{c'}$ denotes the current--current contributions,
i.e.\ the ``tree'' processes in Fig.\ \ref{fig:BdPsiKS}, and the strong
amplitudes $A_{\rm pen}^{q'}$ describe the contributions from penguin 
topologies with internal $q$ quarks ($q\in\{u,c,t\})$. These penguin 
amplitudes take into account both QCD and EW penguin contributions. 
The primes in (\ref{Bd-ampl1}) remind us that we are dealing with a 
$\overline{b}\to\overline{s}$ transition, and the
\begin{equation}\label{lamqs-def}
\lambda_q^{(s)}\equiv V_{qs}V_{qb}^\ast
\end{equation}
are CKM factors. If we employ the unitarity of the CKM matrix
to eliminate $\lambda_t^{(s)}$ through $\lambda_t^{(s)}=-
\lambda_u^{(s)}-\lambda_c^{(s)}$, and the Wolfenstein parametrization, 
we may write
\begin{equation}\label{BdpsiK-ampl2}
A(B_d^0\to J/\psi K_{\rm S})\propto\left[1+\lambda^2 a e^{i\theta}
e^{i\gamma}\right],
\end{equation}
where the hadronic parameter $a e^{i\theta}$ measures, sloppily speaking,
the ratio of penguin- to tree-diagram-like contributions to 
$B_d^0\to J/\psi\,K_{\rm S}$. Since this parameter enters in a
doubly Cabibbo-suppressed way, the formalism discussed in 
Section~\ref{sec:strat} gives, to a very good approximation
\cite{bisa}:
\begin{equation}\label{BpsiK-CP}
{\cal A}_{\rm CP}^{\rm dir}(B_d\to J/\psi K_{\rm S})=0, \quad
{\cal A}_{\rm CP}^{\rm mix}(B_d\to J/\psi K_{\rm S})=-\sin\phi_d.
\end{equation}
After important first steps by the OPAL, CDF and ALEPH collaborations, 
the $B_d\to J/\psi K_{\rm S}$ mode (and similar decays) led eventually, 
in 2001, to the observation of CP violation in the $B$ system 
\cite{BaBar-CP-obs,Belle-CP-obs}. The present status of $\sin2\beta$ is 
given as follows:
\begin{equation}
\sin2\beta=\left\{\begin{array}{ll}
0.741\pm 0.067  \pm0.033  &
\mbox{(BaBar \cite{Babar-s2b-02})}\\
0.719\pm 0.074  \pm0.035  &
\mbox{(Belle \cite{Belle-s2b-02}),}
\end{array}\right.
\end{equation}
yielding the world average \cite{nir} 
\begin{equation}\label{s2b-average}
\sin2\beta=0.734\pm0.054, 
\end{equation}
which agrees well with the results of the ``standard analysis'' of the
unitarity triangle (\ref{SM-ranges}), implying 
$0.6\lsim\sin2\beta\lsim0.9$. 

In the LHC era, the experimental accuracy of the measurement of $\sin2\beta$ 
may be increased by one order of magnitude \cite{LHC-BOOK}. In view of
such a tremendous accuracy, it will then be important to obtain deeper
insights into the theoretical uncertainties affecting (\ref{BpsiK-CP}),
which are due to penguin contributions. A possibility to control them
is provided by the $B_s\to J/\psi K_{\rm S}$ channel \cite{RF-BdsPsiK}. 
Moreover, also direct CP violation in $B\to J/\psi K$ modes allows us
to probe such penguin effects \cite{RF-rev,FM-BpsiK}. So far, there are 
no experimental indications for non-vanishing CP asymmetries of this kind.

Although the agreement between (\ref{s2b-average}) and the results of the
CKM fits is striking, it should not be forgotten that new physics may 
nevertheless hide in ${\cal A}_{\rm CP}^{\rm mix}(B_d\to J/\psi K_{\rm S})$. 
The point is that the key quantity is actually $\phi_d$, which is fixed 
through $\sin\phi_d=0.734\pm0.054$ up to a twofold ambiguity,
\begin{equation}\label{phid-det}
\phi_d=\left(47^{+5}_{-4}\right)^\circ \, \lor \,
\left(133^{+4}_{-5}\right)^\circ.
\end{equation}
Here the former solution would be in perfect agreement with the range
implied by the CKM fits, $40^\circ\lsim\phi_d\lsim60^\circ$, whereas
the latter would correspond to new physics. The two solutions can
be distinguished through a measurement of the sign of $\cos\phi_d$: 
in the case of $\cos\phi_d=+0.7>0$, we would conclude
$\phi_d=47^\circ$, whereas $\cos\phi_d=-0.7<0$ would point towards
$\phi_d=133^\circ$, i.e.\ new physics. 
There are several strategies on the market to resolve the twofold
ambiguity in the extraction of $\phi_d$ \cite{ambig}. Unfortunately,
they are rather challenging from a practical point of view. In the 
$B\to J/\psi K$ system, $\cos\phi_d$ can be extracted from the 
time-dependent angular distribution of the decay products of 
$B_d\to J/\psi[\to\ell^+\ell^-] K^\ast[\to\pi^0K_{\rm S}]$, if the sign 
of a hadronic parameter $\cos\delta$ involving a strong phase $\delta$ 
is fixed through factorization \cite{DDF2,DFN}. Let us note that analyses
of this kind are already in progress at the $B$ factories \cite{itoh}. 

The preferred mechanism for new physics to manifest itself in CP-violating
effects in $B_d\to J/\psi K_{\rm S}$ is through $B^0_d$--$\overline{B^0_d}$
mixing, which arises in the Standard Model from the box diagrams shown
in Fig.~\ref{fig:boxes}. However, new physics may also enter at the 
$B\to J/\psi K$ amplitude level. Employing estimates borrowed from effective 
field theory suggests that the effects are at most ${\cal O}(10\%)$ for 
a generic new-physics scale $\Lambda_{\rm NP}$ in the TeV regime. In order 
to obtain the whole picture, a set of appropriate observables can be 
introduced, using $B_d\to J/\psi K_{\rm S}$ and its charged counterpart 
$B^\pm\to J/\psi K^\pm$ \cite{FM-BpsiK}. So far, these observables do 
not yet indicate any deviation from the Standard Model. 

In the context of new-physics effects in the $B\to J/\psi K$ system, 
it is interesting to note that an upper bound on $\phi_d$ is implied by
an upper bound on $R_b\propto|V_{ub}/V_{cb}|$, as can be seen in
Fig.~\ref{fig:cont-scheme}. To be specific, we have 
\begin{equation}
\sin\beta_{\rm max}=R_b^{\rm max},
\end{equation}
yielding $\left(\phi_d\right)_{\rm max}^{\rm SM}\sim55^\circ$ for 
$R_b^{\rm max}\sim0.46$. As the determination of $R_b$ from semi-leptonic
tree-level decays is very robust concerning the impact of new physics,
$\phi_d\sim 133^\circ$ would require new-physics contributions to 
$B^0_d$--$\overline{B^0_d}$ mixing. As we will see in 
Subsection~\ref{subsec:BsKK}, an interesting connection between the two 
solutions for $\phi_d$ and constraints on $\gamma$ is provided by CP 
violation in $B_d\to\pi^+\pi^-$ \cite{FlMa2}.

\begin{figure}[t]
\begin{center}
\leavevmode
\epsfysize=3.8truecm 
\epsffile{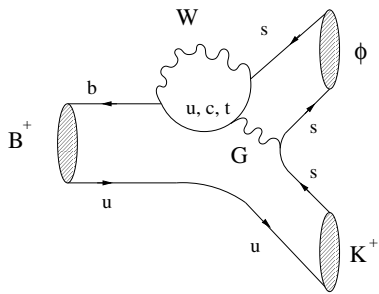} 
\end{center}
\caption{QCD penguin contributions to 
$B^+\to \phi K^+$.}\label{fig:BphiK}
\end{figure}

\subsection{The $B\to\phi K$ System}
An important testing ground for the Standard-Model description of
CP violation is also provided by $B\to \phi K$ decays. As can be 
seen in Fig.~\ref{fig:BphiK}, these modes are governed by QCD penguin 
processes \cite{BphiK-old}, but also EW penguins are sizeable 
\cite{RF-EWP,DH-PhiK}. Consequently, $B\to\phi K$ modes represent a 
sensitive probe for new physics. In the Standard Model, we have the
following relations \cite{RF-rev,growo,loso,FM-BphiK}:
\begin{eqnarray}
{\cal A}_{\rm CP}^{\rm dir}(B_d\to \phi K_{\rm S})&=&0+{\cal O}(\lambda^2)\\
{\cal A}_{\rm CP}^{\rm mix}(B_d\to \phi K_{\rm S})&=&
{\cal A}_{\rm CP}^{\rm mix}(B_d\to J/\psi K_{\rm S})+{\cal O}(\lambda^2).
\end{eqnarray}
As in the case of the $B\to J/\psi K$ system, a combined analysis of
$B_d\to \phi K_{\rm S}$, $B^\pm \to \phi K^\pm$ modes should be 
performed in order to obtain the whole picture \cite{FM-BphiK}. There is 
also the possibility of an unfortunate case, where new physics cannot be
distinguished from the Standard Model, as discussed in detail in 
\cite{RF-Phys-Rep,FM-BphiK}. 

In the summer of 2002, the experimental status can be
summarized as follows:
\begin{equation}
{\cal A}_{\rm CP}^{\rm dir}(B_d\to \phi K_{\rm S})=
\left\{\begin{array}{ll}
\mbox{n.a.} &\mbox{(BaBar \cite{BphiK-BaBar})}\\
0.56\pm0.41\pm0.12
&\mbox{(Belle \cite{BphiK-Belle})}
\end{array}\right.
\end{equation}
\begin{equation}
{\cal A}_{\rm CP}^{\rm mix}(B_d\to \phi K_{\rm S})=
\left\{\begin{array}{ll}
0.19^{+0.50}_{-0.52}\pm 0.09 &\mbox{(BaBar \cite{BphiK-BaBar})}\\
0.73\pm0.64\pm0.18 &\mbox{(Belle \cite{BphiK-Belle}).}
\end{array}\right.
\end{equation}
Unfortunately, the experimental uncertainties are still very large.
Because of ${\cal A}_{\rm CP}^{\rm mix}(B_d\to J/\psi K_{\rm S})=
-0.734 \pm 0.054$ (see (\ref{BpsiK-CP}) and (\ref{s2b-average})), 
there were already speculations about new-physics effects in 
$B_d\to\phi K_{\rm S}$ \cite{BPhiK-NP}. In this context, it is interesting
to note that there are more data available from Belle:
\begin{eqnarray}
{\cal A}_{\rm CP}^{\rm dir}(B_d\to \eta' K_{\rm S})&=&-0.26\pm0.22\pm0.03\\
{\cal A}_{\rm CP}^{\rm mix}(B_d\to \eta' K_{\rm S})
&=&-0.76\pm0.36^{+0.06}_{-0.05}
\end{eqnarray}
\begin{eqnarray}
{\cal A}_{\rm CP}^{\rm dir}(B_d\to K^+K^-K_{\rm S})&=&
0.42\pm0.36\pm0.09^{+0.22}_{-0.03}\\
{\cal A}_{\rm CP}^{\rm mix}(B_d\to K^+K^-K_{\rm S})&=&
-0.52\pm0.46\pm0.11^{+0.03}_{-0.27}.
\end{eqnarray}
The corresponding modes are governed by the same quark-level transitions as
$B_d\to\phi K_{\rm S}$. Consequently, it is probably too early to 
be excited too much by the possibility of signals of new physics in 
$B_d\to\phi K_{\rm S}$ \cite{nir}. However, the experimental situation 
should improve significantly in the future.

\subsection{The Decay $B_d\to\pi^+\pi^-$}
\begin{figure}[t]
\begin{center}
\leavevmode
\epsfysize=3.8truecm 
\epsffile{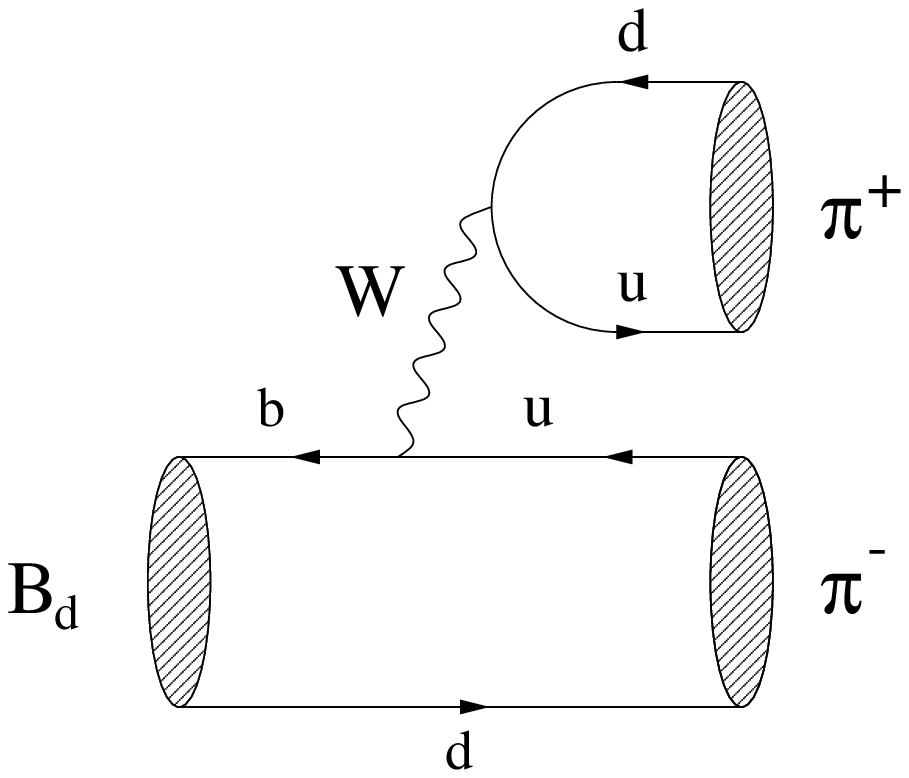} \hspace*{1truecm}
\epsfysize=4.0truecm 
\epsffile{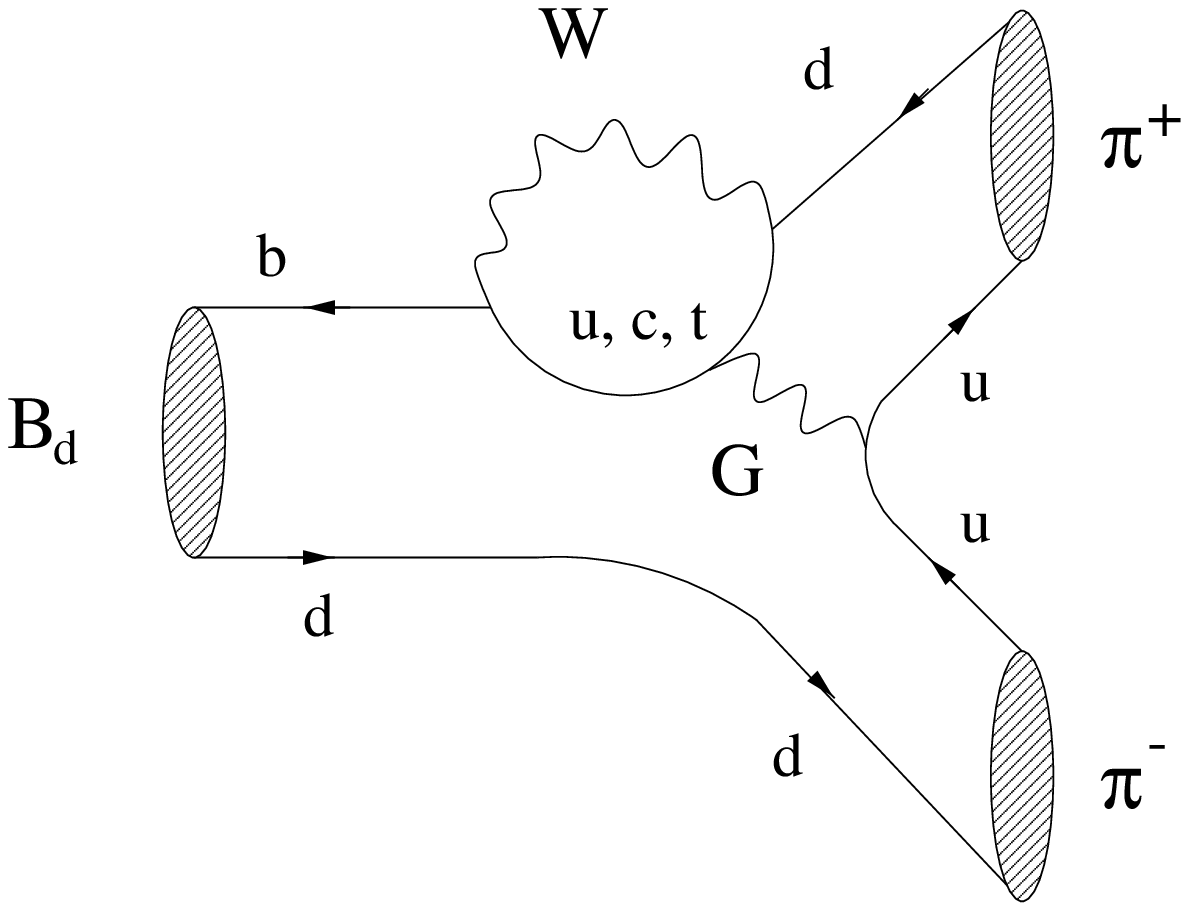}
\end{center}
\caption{Feynman diagrams contributing to 
$B_d^0\to\pi^+\pi^-$.}\label{fig:bpipi}
\end{figure}

Another benchmark mode for the $B$ factories is $B_d^0\to\pi^+\pi^-$,
which is a decay into a CP eigenstate with eigenvalue $+1$, and 
originates from $\overline{b}\to\overline{u} u \overline{d}$ 
quark-level transitions, as can be seen in Fig.~\ref{fig:bpipi}. 
In analogy to (\ref{Bd-ampl1}), the corresponding decay amplitude 
can be written in the following form \cite{RF-BsKK}:
\begin{equation}
A(B_d^0\to\pi^+\pi^-)=
\lambda_u^{(d)}\left(A_{\rm CC}^{u}+
A_{\rm pen}^{u}
\right)+\lambda_c^{(d)}A_{\rm pen}^{c}+\lambda_t^{(d)}A_{\rm pen}^{t}.
\end{equation}
If we use again the unitarity of the CKM matrix, yielding
$\lambda_t^{(d)}=-\lambda_u^{(d)}-\lambda_c^{(d)}$, as well as the
Wolfenstein parametrization, we obtain
\begin{equation}\label{Bpipi-ampl}
A(B_d^0\to\pi^+\pi^-)\propto\left[e^{i\gamma}-de^{i\theta}\right],
\end{equation}
where 
\begin{equation}\label{D-DEF}
d e^{i\theta}\equiv\frac{1}{R_b}
\left(\frac{A_{\rm pen}^{c}-A_{\rm pen}^{t}}{A_{\rm CC}^{u}+
A_{\rm pen}^{u}-A_{\rm pen}^{t}}\right)
\end{equation}
measures, sloppily speaking, the ratio of penguin to tree
contributions in $B_d\to\pi^+\pi^-$. 
In contrast to the $B_d^0\to J/\psi K_{\rm S}$ amplitude
(\ref{BdpsiK-ampl2}), this parameter does {\it not} enter in 
(\ref{Bpipi-ampl}) in a doubly Cabibbo-suppressed way, thereby leading to the 
well-known ``penguin problem'' in $B_d\to\pi^+\pi^-$. If we had 
negligible penguin contributions, i.e.\ $d=0$, the corresponding CP-violating 
observables were given as follows:
\begin{equation}
{\cal A}_{\rm CP}^{\rm dir}(B_d\to\pi^+\pi^-)=0, \quad
{\cal A}_{\rm CP}^{\rm mix}(B_d\to\pi^+\pi^-)=\sin(2\beta+2\gamma)=
-\sin 2\alpha,
\end{equation}
where we have also used the unitarity relation $2\beta+2\gamma=2\pi-2\alpha$. 
We observe that actually the phases $2\beta=\phi_d$ and $\gamma$ 
enter directly in the $B_d\to\pi^+\pi^-$ observables, and not $\alpha$. 
Consequently, since $\phi_d$ can be fixed straightforwardly through 
$B_d\to J/\psi K_{\rm S}$, we may use $B_d\to\pi^+\pi^-$ to probe $\gamma$. 
This is advantageous to deal with penguins and possible new-physics effects, 
as we will see in Subsection~\ref{subsec:BsKK}. 

Measurements of the $B_d\to\pi^+\pi^-$ CP asymmetries are already 
available:
\begin{equation}\label{Adir-exp}
{\cal A}_{\rm CP}^{\rm dir}(B_d\to\pi^+\pi^-)=\left\{
\begin{array}{ll}
-0.30\pm0.25\pm0.04 & \mbox{(BaBar \cite{BaBar-Bpipi})}\\
-0.94^{+0.31}_{-0.25}\pm0.09 & \mbox{(Belle \cite{Belle-Bpipi})}
\end{array}
\right.
\end{equation}
\begin{equation}\label{Amix-exp}
{\cal A}_{\rm CP}^{\rm mix}(B_d\to\pi^+\pi^-)=\left\{
\begin{array}{ll}
-0.02\pm0.34\pm0.05& \mbox{(BaBar \cite{BaBar-Bpipi})}\\
1.21^{+0.27+0.13}_{-0.38-0.16} & \mbox{(Belle \cite{Belle-Bpipi}).}
\end{array}
\right.
\end{equation}
Unfortunately, the BaBar and Belle results are not fully consistent with
each other; the experimental picture will hopefully be clarified soon. 
Forming never-theless the weighted averages of (\ref{Adir-exp}) and 
(\ref{Amix-exp}), using the rules of the Particle Data Group (PDG), yields
\begin{eqnarray}
{\cal A}_{\rm CP}^{\rm dir}(B_d\to\pi^+\pi^-)&=&-0.57\pm0.19 \,\, (0.32)
\label{Bpipi-CP-averages}\\
{\cal A}_{\rm CP}^{\rm mix}(B_d\to\pi^+\pi^-)&=&0.57\pm0.25 \,\, (0.61),
\label{Bpipi-CP-averages2}
\end{eqnarray}
where the errors in brackets are the ones increased by the PDG 
scaling-factor procedure \cite{PDG}. Direct CP violation at this 
level would require large penguin contributions with large CP-conserving 
strong phases. A significant impact of penguins on $B_d\to\pi^+\pi^-$ is 
also indicated by data on $B\to\pi K,\pi\pi$ decays, as well as by 
theoretical considerations \cite{CHARM-PEN,BBNS3,SU,keum} (see 
Subsection~\ref{subsec:BsKK}). Consequently, it is already evident that 
the penguin contributions to $B_d\to\pi^+\pi^-$ {\it cannot} be neglected.

Many approaches to deal with the penguin problem in the extraction 
of weak phases from the CP-violating $B_d\to\pi^+\pi^-$ observables were 
developed; the best known is an isospin analysis of the $B\to\pi\pi$ 
system \cite{GL}, yielding $\alpha$. Unfortunately, this approach is very 
difficult in practice, as it requires a measurement of the 
$B_d^0\to\pi^0\pi^0$ and $\overline{B_d^0}\to\pi^0\pi^0$ branching ratios. 
However, useful bounds 
may already be obtained from experimental constraints on the CP-averaged
$B_d\to\pi^0\pi^0$ branching ratio \cite{alpha-bounds,charles}. 
Alternatively, we may employ the CKM unitarity to express 
${\cal A}_{\rm CP}^{\rm mix}(B_d\to\pi^+\pi^-)$ in terms of $\alpha$ and 
hadronic parameters. Using ${\cal A}_{\rm CP}^{\rm dir}(B_d\to\pi^+\pi^-)$, 
a strong phase can be eliminated, allowing us to determine $\alpha$ as
a function of a hadronic parameter $|p/t|$, which is, however, problematic
to be determined reliably \cite{CHARM-PEN,BBNS1,BBNS3,charles,FM1,LSS,LX}.
A different parametrization of the $B_d\to\pi^+\pi^-$ observables, involving 
a hadronic parameter $P/T$ and $\phi_d=2\beta$, is employed in 
\cite{GR-Bpipi1}, where, moreover, $\alpha+\beta+\gamma=180^\circ$ is used 
to eliminate $\gamma$, and $\beta$ is fixed through the Standard-Model 
solution $\sim 26^\circ$ implied by ${\cal A}_{\rm CP}^{\rm mix}
(B_d\to J/\psi K_{\rm S})$. Provided $|P/T|$ is known, $\alpha$ can be 
extracted. To this end, $SU(3)$ flavour-symmetry arguments and plausible 
dynamical assumptions are 
used to fix $|P|$ through the CP-averaged $B^\pm\to\pi^\pm K$ branching ratio. 
On the other hand, $|T|$ is estimated with the help of factorization and 
data on $B\to\pi\ell\nu$. Refinements of this approach were presented in 
\cite{GR-Bpipi2}. 
Another strategy to deal with penguins in $B_d\to\pi^+\pi^-$ is offered 
by $B_s\to K^+K^-$. Using the $U$-spin flavour symmetry of strong 
interactions, $\phi_d$ and $\gamma$ can be extracted from the corresponding 
CP-violating observables \cite{RF-BsKK}. Before coming back to this approach 
in more detail in Subsection~\ref{subsec:BsKK}, let us first have a closer 
look at the $B_s$-meson system.

\section{The $B_s$-Meson System}\label{sec:Bs-bench}
\subsection{General Features}\label{subsec:Bs-features}
At the $e^+e^-$ $B$ factories operating at the $\Upsilon(4S)$ resonance, 
no $B_s$ mesons are accessible, since $\Upsilon(4S)$ states decay only 
to $B_{u,d}$-mesons, but not to $B_s$. On the other hand, the physics 
potential of the $B_s$ system is very promising for hadron machines, 
where plenty of $B_s$ mesons are produced. Consequently,
$B_s$ physics is in some sense the ``El Dorado'' for $B$ experiments at 
hadron colliders. There are important differences between the $B_d$ and 
$B_s$ systems:
\begin{itemize}
\item Within the Standard Model, the $B^0_s$--$\overline{B^0_s}$ mixing 
phase probes the tiny angle $\delta\gamma$ in the unitarity triangle
shown in Fig.\ \ref{fig:UT} (b), and is hence negligibly small:
\begin{equation}
\phi_s=-2\delta\gamma=-2\lambda^2\eta={\cal O}(-2^\circ),
\end{equation}
whereas $\phi_d=2\beta={\cal O}(50^\circ)$.
\item A large $x_s\equiv\Delta M_s/\Gamma_s={\cal O}(20)$, where
$\Gamma_s\equiv (\Gamma_{\rm H}^{(q)}+\Gamma_{\rm L}^{(q)})/2$, 
is expected in the Standard Model, whereas $x_d=0.775\pm0.012$. The 
present lower bound on $\Delta M_s$ is given as follows \cite{LEPBOSC}:
\begin{equation}\label{Ms-bound}
\Delta M_s>14.4\,\mbox{ps}^{-1} \,\, (95\% \,\,{\rm C.L.}).
\end{equation}
\item There may be a sizeable width difference 
$\Delta\Gamma_s/\Gamma_s={\cal O}(-10\%)$ between the mass eigenstates 
of the $B_s$ system that is due to CKM-favoured $b\to c\overline{c}s$ 
quark-level transitions into final states common to $\overline{B^0_s}$ 
and $B^0_s$, whereas $\Delta\Gamma_d$ is negligibly small \cite{BeLe}.
The present CDF and LEP results imply \cite{LEPBOSC}
\begin{equation}
|\Delta\Gamma_s|/\Gamma_s<0.31 \,\, (95\% \,\,{\rm C.L.}).
\end{equation}
Interesting applications of $\Delta\Gamma_s$ are extractions of weak 
phases from ``untagged'' $B_s$ data samples, where we do not distinguish 
between initially present $B^0_s$ or $\overline{B^0_s}$ mesons, 
as discussed in \cite{Bs-untagged}.
\end{itemize}

\begin{figure}[t]
\centerline{\rotate[r]{
\epsfysize=8.6truecm
{\epsffile{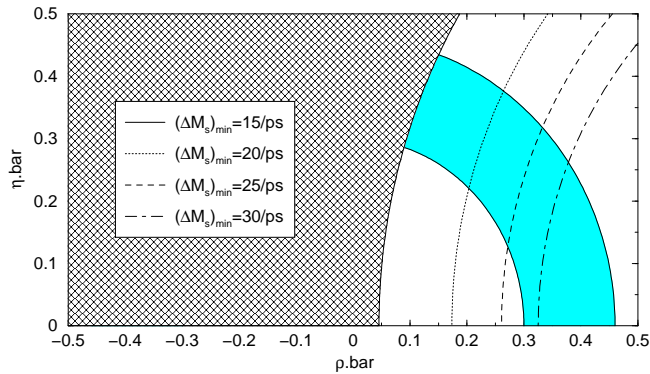}}}}
\caption{The impact of the upper limit $(R_t)_{\rm max}$
on the allowed range in the 
$\overline{\rho}$--$\overline{\eta}$ plane for $\xi=1.15$. The
shaded region corresponds to $R_b=0.38\pm0.08$.}\label{fig:UT-constr}
\end{figure}

Let us now discuss the r\^ole of $\Delta M_s$ for the determination
of the unitarity triangle in more detail. As we have already noted in
Subsection~\ref{subsec:UT-analysis}, the comparison of $\Delta M_d$
with $\Delta M_s$ allows an interesting determination of the side $R_t$ 
of the unitarity triangle. To this end, only a single $SU(3)$-breaking
parameter 
\begin{equation}\label{xi-SU3-def}
\xi\equiv\frac{\sqrt{\hat B_{B_s}f_{B_s}}}{\sqrt{\hat B_{B_d}}f_{B_d}}
=1.15\pm0.06
\end{equation}
is required, which measures $SU(3)$-breaking effects in non-perturbative 
mixing and decay parameters. It can be determined through lattice or 
QCD sum-rule calculations. The mass difference $\Delta M_s$ has not yet 
been measured. However, lower bounds on $\Delta M_s$ can be 
converted into upper bounds on $R_t$ through \cite{Buras-ICHEP96}
\begin{equation}\label{DMs-constr}
\left(R_t\right)_{\rm max}=0.83\times\xi\times\sqrt{\frac{15.0\,
{\rm ps}^{-1}}{\left(\Delta M_s\right)_{\rm min}}},
\end{equation}
excluding already a large part in the $\overline{\rho}$--$\overline{\eta}$ 
plane, as can be seen in Fig.~\ref{fig:UT-constr}. In particular, 
$\gamma<90^\circ$ is implied. In a recent paper \cite{KroRy}, it is argued 
that $\xi$ may actually be significantly larger than the conventional range 
given in (\ref{xi-SU3-def}), $\xi=1.32\pm0.10$ (see also \cite{AS-CKM}). 
In this case, the excluded range in the 
$\overline{\rho}$--$\overline{\eta}$ plane would be reduced, shifting
the upper limit for $\gamma$ closer to $90^\circ$. Hopefully, the status of
$\xi$ will be clarified soon. In the near future, run II of the Tevatron 
should provide a measurement of $\Delta M_s$, thereby constraining the 
unitarity triangle and $\gamma$ in a much more stringent way.

\subsection{Benchmark $B_s$ Decays}
An interesting class of $B_s$ decays is due to $b(\overline{b})\to 
c \overline{u}s(\overline{s})$
quark-level transitions. Here we have to deal with pure ``tree'' decays, 
where both $B_s^0$ and $\overline{B_s^0}$ mesons may decay into the same 
final state $f$. The resulting interference effects between decay and 
mixing processes allow a {\it theoretically clean} extraction of 
$\phi_s+\gamma$ from 
\begin{equation}\label{xi-xi}
\xi_f^{(s)}\times\xi_{\overline{f}}^{(s)}=e^{-2i(\phi_s+\gamma)}.
\end{equation}
There are several well-known strategies on the market employing these 
features: we may consider the colour-allowed decays $B_s\to D_s^\pm K^\mp$ 
\cite{ADK}, or the colour-suppressed modes $B_s\to D^0\phi$ \cite{GL0}. In the
case of $B_s\to D_s^{\ast\pm} K^{\ast\mp}$ or $B_s\to D^{\ast0}\phi$,
the observables of the corresponding angular distributions provide
sufficient information to extract $\phi_s+\gamma$ from ``untagged''
analyses \cite{FD2}, requiring a sizeable $\Delta\Gamma_s$. A
``tagged'' strategy involving $B_s\to D_s^{\ast\pm} K^{\ast\mp}$ modes
was proposed in \cite{LSS-00}. Recently, strategies making use of
``CP-tagged'' $B_s$ decays were proposed \cite{FP}, which require a 
symmetric $e^+e^-$ collider operated at the $\Upsilon(5S)$ resonance. 
In this approach, initially present CP eigenstates $B_s^{\rm CP}$ are 
employed, which can be tagged through the fact that the 
$B_s^0/\overline{B_s^0}$ mixtures have anticorrelated CP eigenvalues 
at $\Upsilon(5S)$. Here $B_s\to D_s^\pm K^\mp, 
D_s^\pm K^{\ast\mp}, D_s^{\ast\pm} K^{\ast\mp}$ modes may be used. 
Let us note that there is also an interesting counterpart of (\ref{xi-xi}) 
in the $B_d$ system \cite{BdDpi}, which employs 
$B_d\to D^{(\ast)\pm}\pi^\mp$ decays, 
and allows a determination of $\phi_d+\gamma$. 

The extraction of $\gamma$ from the phase $\phi_s+\gamma$ provided
by the $B_s$ approaches sketched in the previous paragraph requires 
$\phi_s$ as an additional input, which is negligibly small in the Standard 
Model. Whereas it appears to be quite unlikely that the pure tree decays 
listed above are affected significantly by new physics, as they involve 
no flavour-changing neutral-current processes, this is not the case for 
the $B^0_s$--$\overline{B^0_s}$ mixing phase $\phi_s$. In order to probe 
this quantity, $B_s\to J/\psi\,\phi$ 
offers interesting strategies \cite{DFN,DDF1}. Since this decay is the 
$B_s$ counterpart of $B_d\to J/\psi K_{\rm S}$, the corresponding 
Feynman diagrams are analogous to those shown in Fig.~\ref{fig:BdPsiKS}. 
However, in contrast to $B_d\to J/\psi K_{\rm S}$, the final state of 
$B_s\to J/\psi\phi$ is an admixture of 
different CP eigenstates. In order to disentangle them, we have to use 
the angular distribution of the $J/\psi\to \ell^+\ell^-$ and 
$\phi\to K^+K^-$ decay products~\cite{DDLR}. The corresponding 
observables are governed by \cite{LHC-BOOK}
\begin{equation}\label{Bspsiphi-obs}
\xi^{(s)}_{\psi\phi}\,\propto\, e^{-i\phi_s}
\left[1-2\,i\,\sin\gamma\times{\cal O}(10^{-3})\right].
\end{equation}
Since we have $\phi_s={\cal O}(-2^\circ)$ in the Standard Model, the 
extraction of $\phi_s$ from the $B_s\to J/\psi[\to\ell^+\ell^-] 
\phi[\to K^+K^-]$ angular distribution may well be affected by hadronic 
uncertainties at the $10\%$ level. These hadronic uncertainties, which may 
become an important issue in the LHC era \cite{LHC-BOOK}, can be controlled 
through $B_d\to J/\psi\, \rho^0$, exhibiting some other interesting 
features \cite{RF-ang}. Since $B_s\to J/\psi\phi$ shows small CP-violating 
effects in the Standard Model because of
(\ref{Bspsiphi-obs}), this mode represents a sensitive probe to search for 
new-physics contributions to $B^0_s$--$\overline{B^0_s}$ mixing \cite{NiSi}. 
Note that new-physics effects entering at the $B_s\to J/\psi\phi$ amplitude 
level are expected to play a minor r\^ole and can already be probed in the 
$B\to J/\psi K$ system \cite{FM-BpsiK}. For a detailed discussion of 
``smoking-gun'' signals of a sizeable value of $\phi_s$, see \cite{DFN}. 
There, also methods to determine this phase {\it unambiguously} are proposed.

\begin{figure}[t]
\begin{center}
\leavevmode
\epsfysize=4.0truecm 
\epsffile{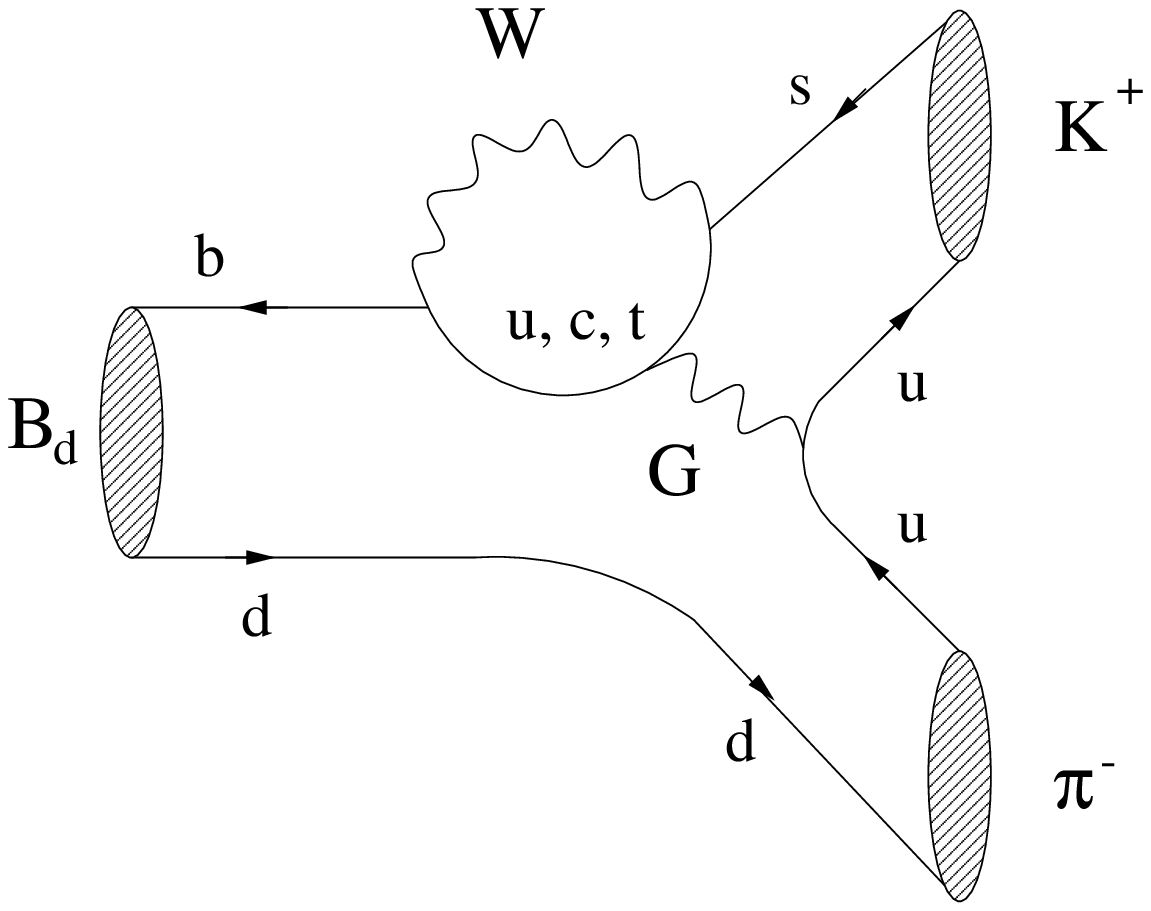} \hspace*{1.4truecm}
\epsfysize=3.5truecm 
\epsffile{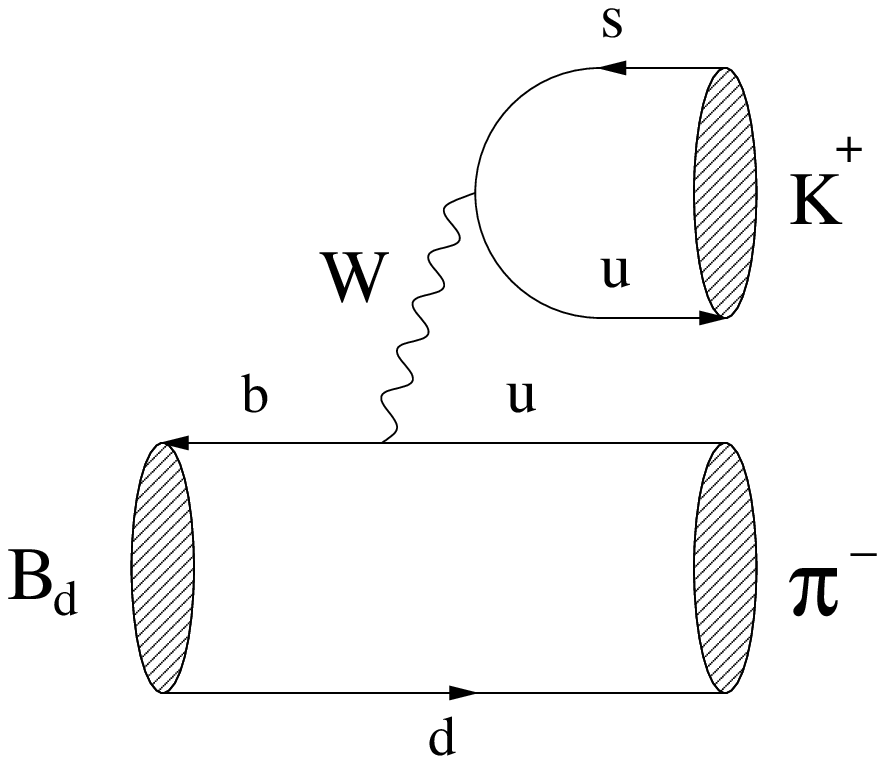}
\end{center}
\vspace*{-0.3truecm}
\caption{Feynman diagrams contributing to 
$B^0_d\to\pi^-K^+$.}\label{fig:BpiK-neutral}
\end{figure}

\section{Recent Developments}\label{sec:recent}
\subsection{Status of $B\to\pi K$ Decays}\label{subsec:BpiK}
If we employ flavour-symmetry arguments and make plausible dynamical 
assumptions, $B\to\pi K$ decays allow determinations of $\gamma$ and 
hadronic parameters with a ``minimal'' theoretical input 
\cite{GRL}--\cite{GR-BpiK-recent}. 
Alternative strategies, relying on a more extensive use of theory, are 
provided by the QCD factorization \cite{BBNS1,BBNS2,BBNS3} and PQCD 
\cite{PQCD,SU} approaches, which furthermore allow a reduction of the 
theoretical uncertainties of the flavour-symmetry strategies. These topics 
are discussed in detail in \cite{neubert}. Let
us here focus on the former kind of strategies. 

To get more familiar with $B\to\pi K$ modes, let us consider 
$B^0_d\to\pi^-K^+$. As can be seen in Fig.~\ref{fig:BpiK-neutral},
this channel receives contributions from penguin and colour-allowed 
tree-diagram-like topologies, where the latter bring $\gamma$ into 
the game. Because of the small ratio 
$|V_{us}V_{ub}^\ast/(V_{ts}V_{tb}^\ast)|\approx0.02$, the QCD penguin
topologies dominate this decay, despite their loop suppression. 
This interesting feature applies to all $B\to\pi K$ modes. Because of the 
large top-quark mass, we also have to care about EW penguins. However, in 
the case of $B^0_d\to\pi^-K^+$ and $B^+\to\pi^+K^0$, these topologies 
contribute only in colour-suppressed form and are hence expected to play 
a minor r\^ole. On the other hand, EW penguins contribute also in 
colour-allowed form to $B^+\to\pi^0K^+$ and $B^0_d\to\pi^0K^0$, and may 
here even compete with tree-diagram-like topologies. Because of the 
dominance of penguin topologies, $B\to\pi K$ modes are sensitive probes 
for new-physics effects \cite{BPIK-NP}.

\begin{table}[t]
\centering
\begin{tabular}{|c||c|c|c|c|}
\hline
Observable & CLEO \cite{CLEO-BpiK}  & BaBar \cite{BaBar-Bpipi,BaBar-BpiK}  & 
Belle \cite{Belle-BpiK} & Average\\
\hline
$R$ & $1.00\pm0.30$ & $1.08\pm0.15$ & $1.22\pm0.26$ & $1.10\pm0.12$\\
$R_{\rm c}$ & $1.27\pm0.47$ & $1.46\pm0.25$ & $1.34\pm0.37$ & $1.40\pm0.19$\\
$R_{\rm n}$ & $0.59\pm0.27$ & $0.86\pm0.15$ & $1.41\pm0.65$ & $0.82\pm0.13$\\
\hline
\end{tabular}
\caption{CP-conserving $B\to \pi K$ observables as defined in
(\ref{mixed-obs})--(\ref{neut-obs}). For the evaluation of $R$, we have
used $\tau_{B^+}/\tau_{B^0_d}=1.060\pm0.029$.}\label{tab:BPIK-obs}
\end{table}

\begin{table}[t]
\centering
\begin{tabular}{|c||c|c|c|c|}
\hline
Observable & CLEO \cite{CLEO-BpiK-CPV} & BaBar \cite{BaBar-Bpipi,BaBar-BpiK} 
& Belle \cite{Belle-BpiK} & Average\\
\hline
$A_0$ & $0.04\pm0.16$ & $0.11\pm0.06$ & $0.07\pm0.11$ & $0.09\pm0.05$\\
$A_0^{\rm c}$ & $0.37\pm0.32$ & $0.13\pm0.13$ & $0.03\pm0.26$ & $0.14\pm0.11$\\
$A_0^{\rm n}$ & $0.02\pm0.10$ & $0.09\pm0.05$ & $0.08\pm0.13$ & $0.08\pm0.04$\\
\hline
\end{tabular}
\caption{CP-violating $B\to \pi K$ observables as defined in
(\ref{mixed-obs})--(\ref{neut-obs}). For the evaluation
of $A_0$, we have used 
$\tau_{B^+}/\tau_{B^0_d}=1.060\pm0.029$.}\label{tab:BPIK-obs-CPV}
\end{table}

Relations between the $B\to\pi K$ amplitudes that are implied by the $SU(2)$
isospin flavour symmetry of strong interactions suggest the following 
combinations to probe $\gamma$: the ``mixed'' $B^\pm\to\pi^\pm K$, 
$B_d\to\pi^\mp K^\pm$ system \cite{PAPIII}--\cite{defan}, the ``charged'' 
$B^\pm\to\pi^\pm K$, $B^\pm\to\pi^0K^\pm$ system 
\cite{NR}--\cite{BF-neutral1}, and the ``neutral'' $B_d\to\pi^0 K$, 
$B_d\to\pi^\mp K^\pm$ system \cite{BF-neutral1,BF-neutral2}. 
Correspondingly, we may introduce the following sets of observables 
\cite{BF-neutral1}:
\begin{equation}\label{mixed-obs}
\mbox{}\hspace*{0.4truecm}\left\{\begin{array}{c}R\\A_0\end{array}\right\}
\equiv\left[\frac{\mbox{BR}(B^0_d\to\pi^-K^+)\pm
\mbox{BR}(\overline{B^0_d}\to\pi^+K^-)}{\mbox{BR}(B^+\to\pi^+K^0)+
\mbox{BR}(B^-\to\pi^-\overline{K^0})}\right]\frac{\tau_{B^+}}{\tau_{B^0_d}}
\end{equation}
\begin{equation}\label{charged-obs}
\left\{\begin{array}{c}R_{\rm c}\\A_0^{\rm c}\end{array}\right\}
\equiv2\left[\frac{\mbox{BR}(B^+\to\pi^0K^+)\pm
\mbox{BR}(B^-\to\pi^0K^-)}{\mbox{BR}(B^+\to\pi^+K^0)+
\mbox{BR}(B^-\to\pi^-\overline{K^0})}\right]
\end{equation}
\begin{equation}\label{neut-obs}
\left\{\begin{array}{c}R_{\rm n}\\A_0^{\rm n}\end{array}\right\}
\equiv\frac{1}{2}\left[\frac{\mbox{BR}(B^0_d\to\pi^-K^+)\pm
\mbox{BR}(\overline{B^0_d}\to\pi^+K^-)}{\mbox{BR}(B^0_d\to\pi^0K^0)+
\mbox{BR}(\overline{B^0_d}\to\pi^0\overline{K^0})}\right].
\end{equation}
The experimental status of these observables is summarized in 
Tables~\ref{tab:BPIK-obs} and \ref{tab:BPIK-obs-CPV}.
Moreover, there are stringent constraints on CP violation in 
$B^\pm\to\pi^\pm K$:
\begin{equation}\label{ACP-Bc}
{\cal A}_{\rm CP}(B^\pm\to\pi^\pm K)=\left\{\begin{array}{ll}
0.17\pm0.10\pm0.02 & \mbox{(BaBar \cite{BaBar-BpiK})}\\ 
-0.46\pm0.15\pm0.02 & \mbox{(Belle \cite{Belle-BpiK}).}
\end{array}\right.
\end{equation}
Let us note that a very recent preliminary study of Belle indicates that
the large asymmetry in (\ref{ACP-Bc}) is due to a $3 \sigma$ fluctuation
\cite{suzuki}. Within the Standard Model, a sizeable value of 
${\cal A}_{\rm CP}(B^\pm\to\pi^\pm K)$ could be induced by large 
rescattering effects. Other important indicators for
such processes are branching ratios for $B\to KK$ decays, which are already 
strongly constrained by the $B$ factories, and would allow us to take
into account rescattering effects in the extraction of $\gamma$ from
$B\to\pi K$ modes \cite{defan,neubert-BpiK,BF-neutral1,FSI-incl}. Let us note 
that also the QCD factorization approach 
\cite{neubert,BBNS1,BBNS2,BBNS3} is not in favour of large
rescattering processes. For simplicity, we shall neglect such effects in 
the discussion given below. Interestingly, already CP-averaged $B\to\pi K$ 
branching ratios may lead to non-trivial constraints on $\gamma$ 
\cite{FM,NR,BF-neutral1}, provided the corresponding $R_{\rm (c,n)}$ 
observables are found to be sufficiently different from 1. The final 
goal is, however, to determine $\gamma$.

\begin{figure}[t]
\vspace*{-0.1cm}
$$\hspace*{-1.cm}
\epsfysize=0.16\textheight
\epsfxsize=0.26\textheight
\epsffile{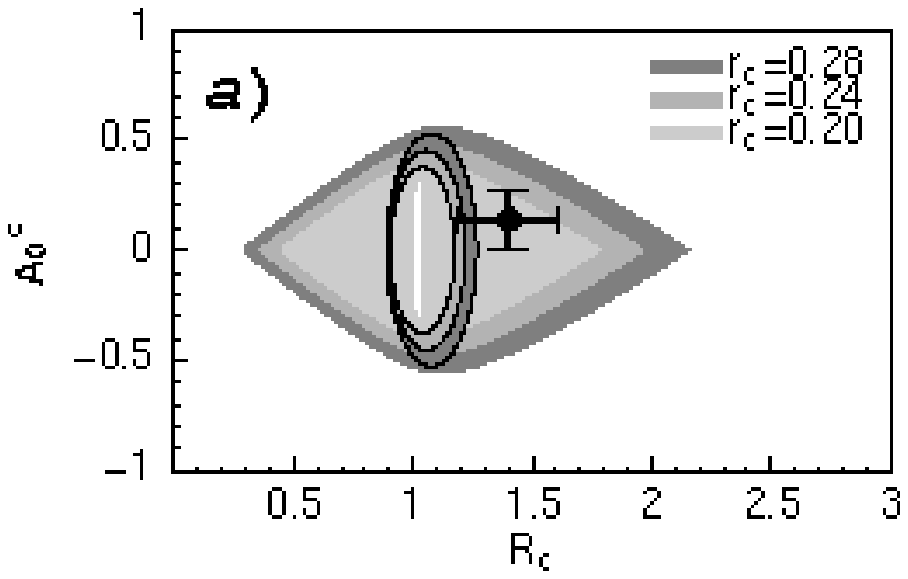} \hspace*{0.3cm}
\epsfysize=0.16\textheight
\epsfxsize=0.26\textheight
\epsffile{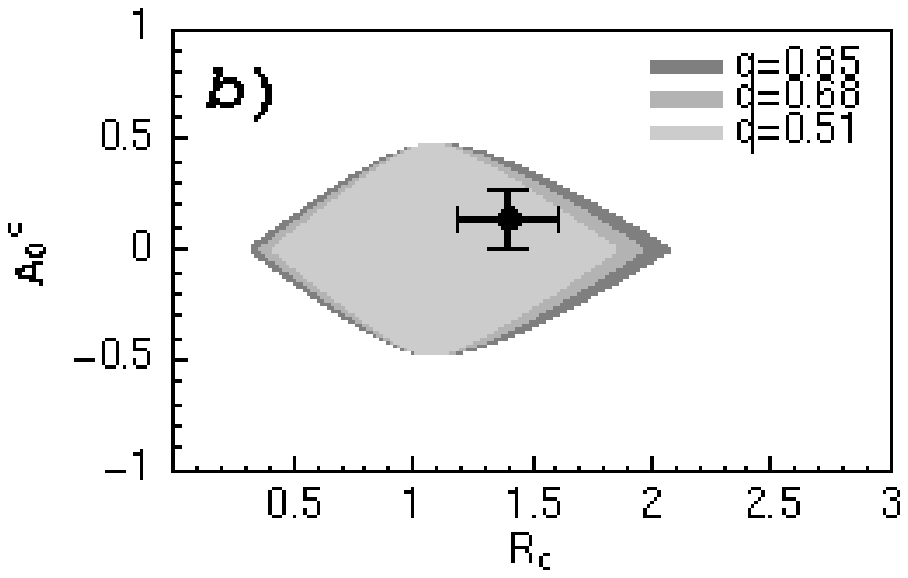}
$$
\vspace*{-0.3cm}
$$\hspace*{-1.cm}
\epsfysize=0.16\textheight
\epsfxsize=0.26\textheight
\epsffile{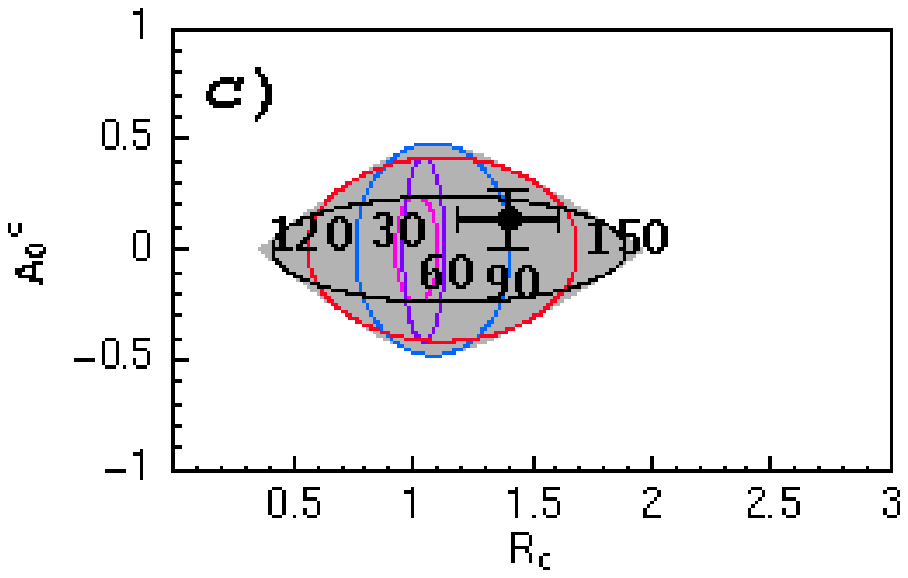} \hspace*{0.3cm}
\epsfysize=0.16\textheight
\epsfxsize=0.26\textheight
\epsffile{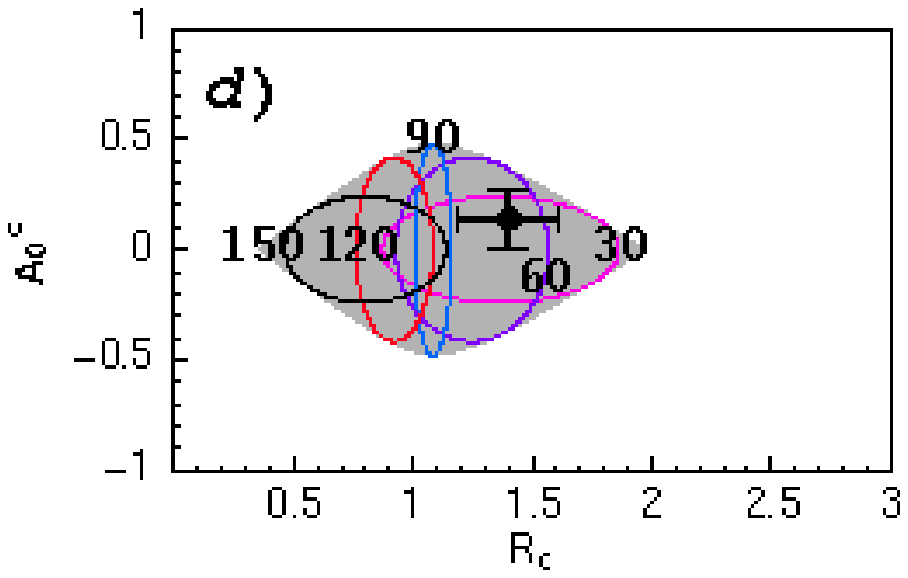}
$$
\caption[]{The allowed regions in the $R_{\rm c}$--$A_0^{\rm c}$ plane: (a) 
corresponds to $0.20 \leq r_{\rm c} \leq 0.28$ 
for $q=0.68$, and (b) to $0.51 \leq q \leq 0.85$ for $r_{\rm c}=0.24$;
the elliptical regions arise if we restrict $\gamma$ to the Standard-Model
range specified in (\ref{gamma-SM}). In (c) and (d), we show also the contours 
for fixed values of $\gamma$ and $|\delta_{\rm c}|$, respectively 
($r_{\rm c}=0.24$, $q=0.68$).}\label{fig:BpiK-charged}
\end{figure}

Let us first turn to the charged and neutral $B\to\pi K$ systems in more
detail. The starting point of our considerations are relations between the 
charged and neutral $B\to\pi K$ amplitudes that follow from the $SU(2)$ 
isospin symmetry of strong interactions. Assuming moreover that the 
rescattering effects discussed above are small, we arrive at a 
parametrization of the following structure \cite{BF-neutral1} 
(for an alternative one, see \cite{neubert-BpiK}):
\begin{eqnarray}
R_{\rm c,n}&=&1-2r_{\rm c,n}\left(\cos\gamma-q\right)\cos\delta_{\rm c,n}
+\left(1-2q\cos\gamma+q^2\right)r_{\rm c,n}^2\label{Rcn-par}\\
A_0^{\rm c,n}&=&2r_{\rm c,n}\sin\delta_{\rm c,n}\sin\gamma.\label{Acn-par}
\end{eqnarray}
Here $r_{\rm c,n}$ measures -- simply speaking -- the ratio of tree to 
penguin topologies. Using $SU(3)$ flavour-symmetry arguments and data on the 
CP-averaged $B^\pm\to\pi^\pm\pi^0$ branching ratio \cite{GRL}, we obtain 
$r_{\rm c,n}\sim0.2$. The parameter $q$ describes the ratio of EW penguin 
to tree contributions, and can be fixed through $SU(3)$ flavour-symmetry 
arguments, yielding $q\sim 0.7$ \cite{NR}. In order to simplify 
(\ref{Rcn-par}) and (\ref{Acn-par}), we have assumed that $q$ is a real 
parameter, as is the case in the strict $SU(3)$ limit; for generalizations, 
see \cite{BF-neutral1}. Finally, $\delta_{\rm c,n}$ is the CP-conserving
strong phase between trees and penguins. 
Consequently, the observables $R_{\rm c,n}$ and $A_0^{\rm c,n}$ depend on 
the two ``unknowns'' $\delta_{\rm c,n}$ and $\gamma$. If we vary them within 
their allowed ranges, i.e.\ $-180^\circ\leq \delta_{\rm c,n}\leq+180^\circ$ 
and $0^\circ\leq \gamma \leq180^\circ$, we obtain an allowed region in 
the $R_{\rm c,n}$--$A_0^{\rm c,n}$ plane \cite{FlMa2,FlMa1}. Should the
measured values of $R_{\rm c,n}$ and $A_0^{\rm c,n}$ lie outside
this region, we would have an immediate signal for new physics. On the
other hand, should the measurements fall into the allowed range,
$\gamma$ and $\delta_{\rm c,n}$ could be extracted. In this case, $\gamma$
could be compared with the results of alternative strategies and with
the values implied by the ``standard analysis'' of the unitarity triangle
discussed in Subsection~\ref{subsec:UT-analysis}, whereas 
$\delta_{\rm c,n}$ provides valuable insights into hadron dynamics, 
thereby allowing tests of theoretical predictions.

\begin{figure}[t]
$$\hspace*{-1.cm}
\epsfysize=0.17\textheight
\epsfxsize=0.27\textheight
\epsffile{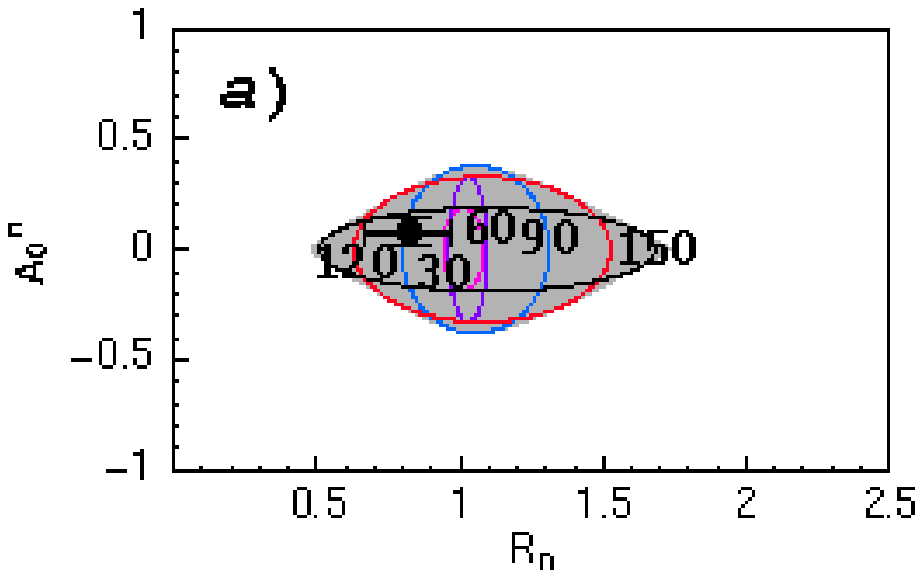} \hspace*{0.3cm}
\epsfysize=0.17\textheight
\epsfxsize=0.27\textheight
\epsffile{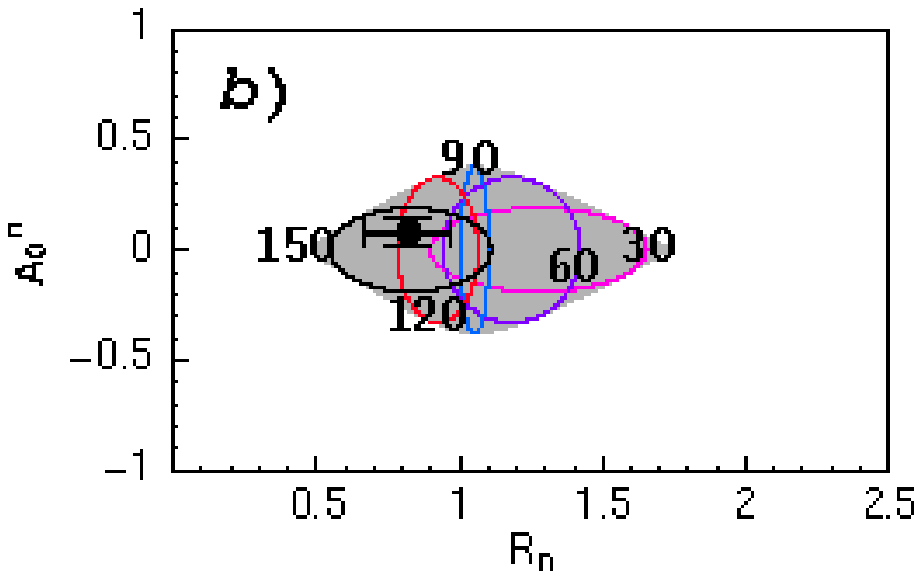}
$$
\caption[]{The allowed regions in the $R_{\rm n}$--$A_0^{\rm n}$ plane
for $q=0.68$ and $r_{\rm n}=0.19$. In (a) and (b), we show also the 
contours for fixed values of $\gamma$ and $|\delta_{\rm n}|$, 
respectively.}\label{fig:BpiK-neutral-cont}
\end{figure}

In Fig.~\ref{fig:BpiK-charged}, we show the allowed regions in the 
$R_{\rm c}$--$A_0^{\rm c}$ plane for various parameter sets 
\cite{FlMa2}. The crosses represent the averages of the experimental 
results given in Tables~\ref{tab:BPIK-obs} and \ref{tab:BPIK-obs-CPV}. 
If $\gamma$ is constrained to the range implied by the ``standard analysis'' 
of the unitarity triangle,
\begin{equation}\label{gamma-SM}
50^\circ\lsim\gamma\lsim70^\circ,
\end{equation}
a much more restricted region arises in the 
$R_{\rm c}$--$A_0^{\rm c}$ plane. The contours in 
Figs.~\ref{fig:BpiK-charged} (c) and (d) allow us to read off 
easily the preferred values for $\gamma$ and $\delta_{\rm c}$,
respectively, from the measured observables \cite{FlMa2}. Interestingly,
the present data seem to favour $\gamma\gsim90^\circ$, which would be in 
conflict with (\ref{gamma-SM}). Moreover, they point towards 
$|\delta_{\rm c}|\lsim90^\circ$; factorization predicts 
$\delta_{\rm c}$ to be close to $0^\circ$ \cite{BBNS3}. The situation
for the neutral $B\to\pi K$ system is illustrated in 
Fig.~\ref{fig:BpiK-neutral-cont}. Interestingly, here the data 
point to $\gamma\gsim90^\circ$ as well, but favour also 
$|\delta_{\rm n}|\gsim90^\circ$ because of the average of 
$R_{\rm n}$ being smaller than 1 \cite{FlMa2,BF-neutral2}. However, 
as can be seen in Table~\ref{tab:BPIK-obs}, the present data are
unfortunately rather unsatisfactory in this respect. 

If future, more accurate data  
really yield a value for $\gamma$ in the second quadrant, the discrepancy 
with (\ref{gamma-SM}) may be due to new-physics contributions to 
$B^0_q$--$\overline{B^0_q}$ mixing ($q\in\{d,s\}$), or to the $B\to\pi K$ 
decay amplitudes. In the former case, the constraints implied by
(\ref{DMs-constr}), which rely on the Standard-Model interpretation of 
$B^0_q$--$\overline{B^0_q}$ mixing, would no longer hold, so that 
$\gamma$ may actually be larger than $90^\circ$. In the latter 
case, the Standard-Model expressions (\ref{Rcn-par}) and (\ref{Acn-par}) 
would receive corrections due to the presence of new physics, so that 
also the extracted value for $\gamma$ would not correspond to the 
Standard-Model result. In such a scenario -- an example would be given by 
new-physics contributions to the EW penguin sector -- also the extracted 
values for $\delta_{\rm c}$ and $\delta_{\rm n}$ may actually no longer 
satisfy $\delta_{\rm c}\approx\delta_{\rm n}$ \cite{BF-neutral2}.

An analysis similar to the one discussed above can also be performed
for the mixed $B\to\pi K$ system, consisting of $B^\pm\to\pi^\pm K$, 
$B_d\to\pi^\mp K^\pm$ modes. To this end, only straightforward 
replacements of variables have to be made. The present data fall 
well into the Standard-Model region in observable space, but do not yet 
allow us to draw further definite conclusions \cite{FlMa2}. At present, 
the situation in the charged and neutral $B\to\pi K$ systems appears to 
be more exciting.

There are also many other recent analyses of $B\to\pi K$ modes. 
For example, a study complementary to the one in $B\to\pi K$
observable space was performed in \cite{ital-corr}, where the allowed 
regions in the $\gamma$--$\delta_{\rm c,n}$ planes implied 
by $B\to\pi K$ data were explored. Another recent $B\to\pi K$ analysis 
can be found in \cite{GR-BpiK-recent}, where the $R_{\rm (c)}$ were 
calculated for given values of $A_0^{\rm (c)}$ as functions of $\gamma$, 
and were compared with the $B$-factory data. Making more extensive use
of theory than in the flavour-symmetry strategies discussed above,
several different avenues to extract $\gamma$ from $B\to\pi K$ modes 
are provided by the QCD factorization approach \cite{neubert,BBNS3}, 
which allows also a reduction of the theoretical uncertainties of the 
flavour-symmetry approaches discussed above, in particular a better control 
of $SU(3)$-breaking effects. In order to analyse $B\to\pi K$ data, also 
sum rules relating CP-averaged branching ratios and CP asymmetries of 
$B \to \pi K$ modes may be useful \cite{matias}.

\subsection{The $B_d\to\pi^+\pi^-$, $B_s\to K^+K^-$ 
System}\label{subsec:BsKK}
As can be seen from Fig.~\ref{fig:bpipi}, $B_d\to\pi^+\pi^-$ is
related to $B_s\to K^+K^-$ through an interchange of all down and 
strange quarks. The corresponding decay amplitudes can be expressed
as follows \cite{RF-BsKK}:
\begin{equation}\label{Bpipi-ampl2}
A(B_d^0\to\pi^+\pi^-)={\cal C}\left[e^{i\gamma}-d e^{i\theta}\right]
\end{equation}
\begin{equation}\label{BsKK-ampl}
A(B_s^0\to K^+K^-)=\left(\frac{\lambda}{1-\lambda^2/2}\right){\cal C}'
\left[e^{i\gamma}+\left(\frac{1-\lambda^2}{\lambda^2}\right)
d'e^{i\theta'}\right],
\end{equation}
where $de^{i\theta}$ was already introduced in (\ref{D-DEF}), 
$d'e^{i\theta'}$ is its $B_s\to K^+K^-$ counterpart, and ${\cal C}$,
${\cal C}'$ are CP-conserving strong amplitudes. Using these general 
parametrizations, we obtain
\begin{equation}\label{Bpipi-obs}
{\cal A}_{\rm CP}^{\rm dir}(B_d\to\pi^+\pi^-)=
\mbox{fct}(d,\theta,\gamma), \quad
{\cal A}_{\rm CP}^{\rm mix}(B_d\to\pi^+\pi^-)=
\mbox{fct}(d,\theta,\gamma,\phi_d)
\end{equation}
\begin{equation}\label{BsKK-obs}
{\cal A}_{\rm CP}^{\rm dir}(B_s\to K^+K^-)=
\mbox{fct}(d',\theta',\gamma), \quad
{\cal A}_{\rm CP}^{\rm mix}(B_s\to K^+K^-)=
\mbox{fct}(d',\theta',\gamma,\phi_s),
\end{equation}
where $\phi_s$ is negligibly small in the Standard Model, or can be fixed
through $B_s\to J/\psi \phi$.  We have 
hence four observables at our disposal, depending on six 
``unknowns''. However, since $B_d\to\pi^+\pi^-$ and $B_s\to K^+K^-$ are 
related to each other by interchanging all down and strange quarks, the 
$U$-spin flavour symmetry of strong interactions implies
\begin{equation}\label{U-spin-rel}
d'e^{i\theta'}=d\,e^{i\theta}.
\end{equation}
Using this relation, the four observables in (\ref{Bpipi-obs}) and
(\ref{BsKK-obs}) depend on the four quantities $d$, $\theta$, 
$\phi_d$ and $\gamma$, which can hence be determined \cite{RF-BsKK}. 
The theoretical accuracy is only limited by the $U$-spin symmetry, 
as no dynamical assumptions about rescattering processes have to be made. 
Theoretical considerations give us confidence in (\ref{U-spin-rel}), since
this relation does not receive $U$-spin-breaking corrections within the
factorization approach \cite{RF-BsKK}. Moreover, we may also obtain 
experimental insights into $U$-spin breaking \cite{RF-BsKK,gronau-U-spin}. 
The $U$-spin arguments can be minimized, if the $B^0_d$--$\overline{B^0_d}$ 
mixing phase $\phi_d$, which can be fixed through $B_d\to J/\psi K_{\rm S}$, 
is used as an input. We may then determine $\gamma$, as well as the hadronic 
quantities $d$, $\theta$, $\theta'$, by using only the $U$-spin relation 
$d'=d$; for a detailed illustration, see \cite{RF-BsKK}. This approach 
is very promising for run II of the Tevatron and the experiments of the LHC 
era, where experimental accuracies for $\gamma$ of ${\cal O}(10^\circ)$ 
\cite{TEV-BOOK} and ${\cal O}(1^\circ)$ \cite{LHC-BOOK} may be achieved, 
respectively. For other recently developed $U$-spin strategies, the reader 
is referred to \cite{RF-BdsPsiK,RF-ang,GR-Uspin,skands}.

\begin{figure}[t]
\centerline{{
\vspace*{-0.2truecm}
\epsfysize=4.4truecm
{\epsffile{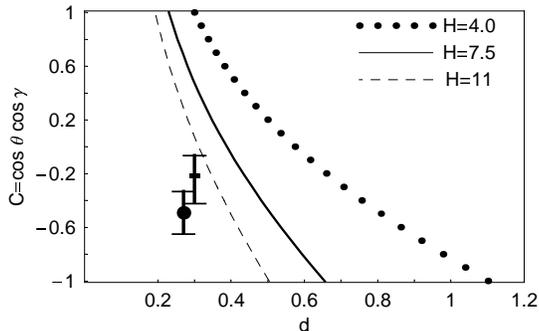}}}}
\caption{Dependence of $C\equiv\cos\theta\cos\gamma$ on $d$ for values of 
$H$ consistent with (\ref{H-det}). The ``circle'' and ``square'' with error 
bars represent the predictions of QCD factorization \cite{BBNS3} and 
PQCD \cite{SU}, respectively, for the Standard-Model range (\ref{gamma-SM}) 
of $\gamma$.}\label{fig:C-d}
\end{figure}

Since $B_s\to K^+K^-$ is not accessible at the $e^+e^-$ $B$ factories 
operating at $\Upsilon(4S)$, data are not yet available. However, as 
can be seen by looking at the corresponding Feynman diagrams, 
$B_s\to K^+K^-$ is related to $B_d\to\pi^\mp K^\pm$ through an 
interchange of spectator quarks. Consequently, we have
\begin{equation}\label{CP-asym-rel}
{\cal A}_{\rm CP}^{\rm dir}(B_s\to K^+K^-)\approx{\cal A}_{\rm CP}^{\rm dir}
(B_d\to\pi^\mp K^\pm)
\end{equation}
\begin{equation}\label{BR-rel}
\mbox{BR}(B_s\to K^+K^-)
\approx\mbox{BR}(B_d\to\pi^\mp K^\pm)\,\frac{\tau_{B_s}}{\tau_{B_d}}.
\end{equation}
For the following considerations, the quantity 
\begin{equation}\label{H-def}
H\equiv\frac{1}{\epsilon}\left|
\frac{{\cal C}'}{{\cal C}}\right|^2
\left[\frac{\mbox{BR}(B_d\to\pi^+\pi^-)}{\mbox{BR}(B_s\to K^+K^-)}\right]
\end{equation}
is particularly useful \cite{U-variant}, where 
$\epsilon\equiv\lambda^2/(1-\lambda^2)$. Using (\ref{BR-rel}), as well as 
factorization to estimate $U$-spin-breaking corrections to 
$|{\cal C}'|=|{\cal C}|$, $H$ can be determined from the $B$-factory data 
as follows:
\begin{equation}\label{H-det}
H\approx\frac{1}{\epsilon}\left(\frac{f_K}{f_\pi}\right)^2
\left[\frac{\mbox{BR}(B_d\to\pi^+\pi^-)}{\mbox{BR}(B_d\to\pi^\mp K^\pm)}
\right]=
\left\{\begin{array}{ll}
7.3\pm2.9 & \mbox{(CLEO \cite{CLEO-BpiK})}\\
7.6\pm1.2 & \mbox{(BaBar \cite{BaBar-BpiK})}\\
7.1\pm1.9 & \mbox{(Belle \cite{Belle-BpiK}).}
\end{array}\right.
\end{equation}
If we employ the $U$-spin relation (\ref{U-spin-rel}) and the amplitude 
parametrizations in (\ref{Bpipi-ampl2}) and (\ref{BsKK-ampl}), we obtain
\begin{equation}\label{H-expr}
H=\frac{1-2 d\cos\theta\cos\gamma+d^2}{\epsilon^2+
2\epsilon d\cos\theta\cos\gamma+d^2}.
\end{equation}
Consequently, $H$ allows us to determine $C\equiv\cos\theta\cos\gamma$ 
as a function of $d$, as shown in Fig.~\ref{fig:C-d}. We observe that
the data imply the rather restricted range $0.2\lsim d\lsim 1$, thereby
indicating that penguins cannot be neglected in $B_d\to\pi^+\pi^-$
analyses. Moreover, the experimental curves are not in favour of a 
Standard-Model interpretation of the theoretical predictions for 
$de^{i\theta}$ obtained within the QCD factorization \cite{BBNS3} and 
PQCD \cite{SU} approaches. Interestingly, agreement could easily be 
achieved for $\gamma>90^\circ$, as the circle and square in
Fig.~\ref{fig:C-d}, calculated for $\gamma=60^\circ$, would then move 
to positive values of $C$ \cite{FlMa2,U-variant}. 

Let us now come back to the decay $B_d\to\pi^+\pi^-$ and its CP-violating
observables, as parametrized in (\ref{Bpipi-obs}). As we have already noted,
$\phi_d$ entering ${\cal A}_{\rm CP}^{\rm mix}(B_d\to\pi^+\pi^-)$
can be fixed through ${\cal A}_{\rm CP}^{\rm mix}(B_d\to J/\psi K_{\rm S})$,
yielding the twofold solution in (\ref{phid-det}). In order to deal with
the penguins, we may employ $H$ as an additional observable. Applying
(\ref{U-spin-rel}), we obtain $H=\mbox{fct}(d,\theta,\gamma)$ (see
(\ref{H-expr})). We may then eliminate $d$ in (\ref{Bpipi-obs}) through 
$H$. If we vary the remaining parameters $\theta$ and $\gamma$ within 
their physical ranges, i.e.\ $-180^\circ\leq \theta\leq+180^\circ$ and 
$0^\circ\leq \gamma \leq180^\circ$, we obtain an allowed region in the
${\cal A}_{\rm CP}^{\rm dir}(B_d\to\pi^+\pi^-)$--${\cal A}_{\rm CP}^{\rm 
mix}(B_d\to\pi^+\pi^-)$ plane.

\begin{figure}[t]
$$\hspace*{-1.cm}
\epsfysize=0.2\textheight
\epsfxsize=0.3\textheight
\epsffile{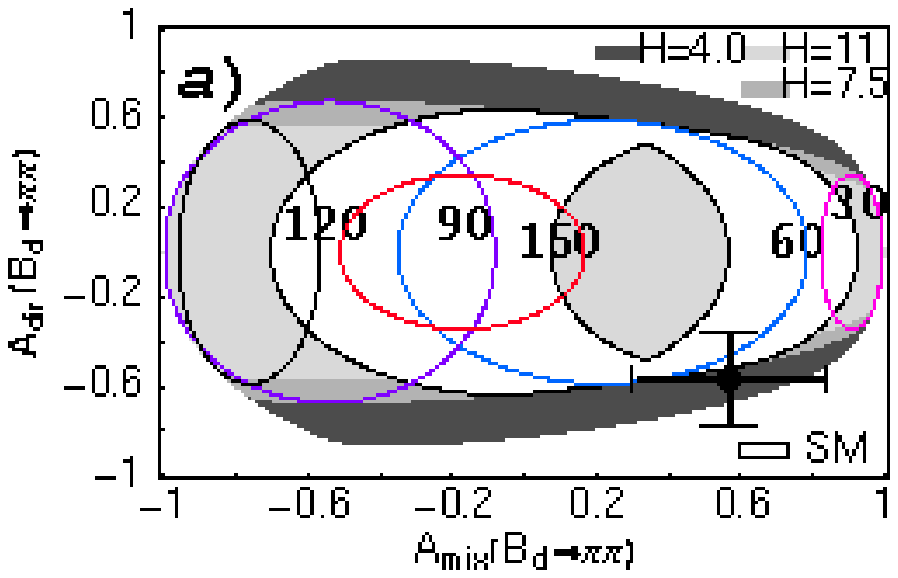} \hspace*{0.3cm}
\epsfysize=0.2\textheight
\epsfxsize=0.3\textheight
\epsffile{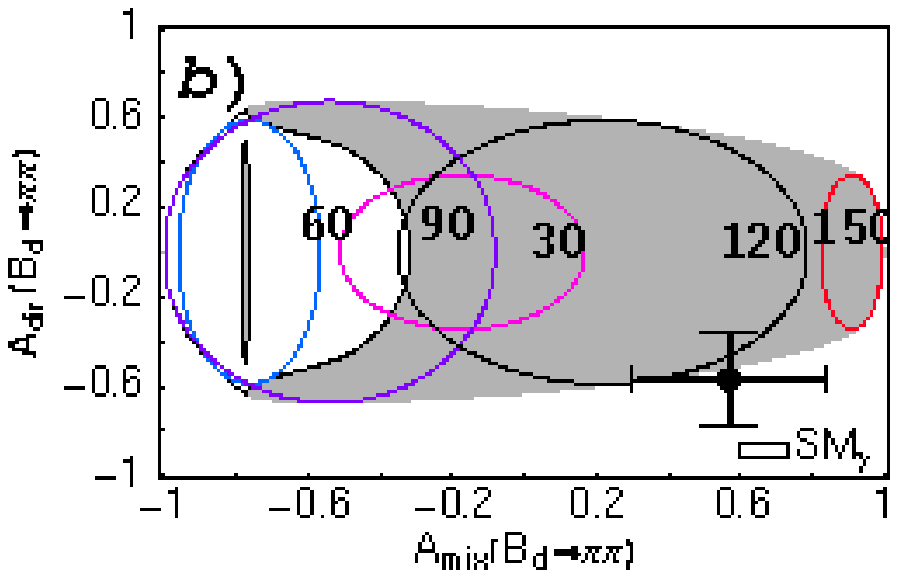}
$$
\caption[]{Allowed regions in the 
${\cal A}_{\rm CP}^{\rm mix}(B_d\to\pi^+\pi^-)$--${\cal A}_{\rm CP}^{\rm
dir}(B_d\to\pi^+\pi^-)$ plane for (a) $\phi_d=47^\circ$ and various values
of $H$, and (b) $\phi_d=133^\circ$ ($H=7.5$). The SM regions arise if we 
restrict $\gamma$ to (\ref{gamma-SM}). Contours representing
fixed values of $\gamma$ are also included.}\label{fig:AdAmpipi}
\end{figure}

\begin{figure}[t]
$$\hspace*{-1.cm}
\epsfysize=0.19\textheight
\epsfxsize=0.29\textheight
\epsffile{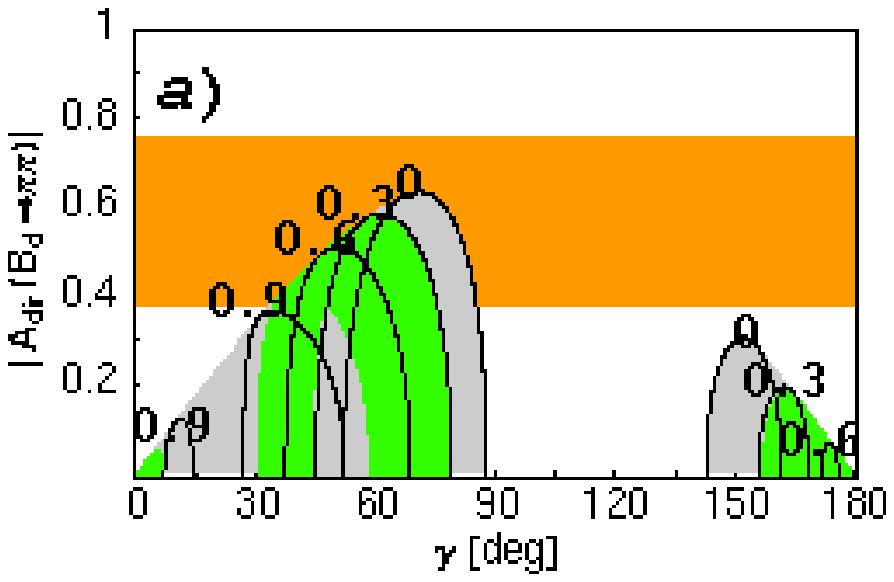} \hspace*{0.3cm}
\epsfysize=0.2\textheight
\epsfxsize=0.3\textheight
\epsffile{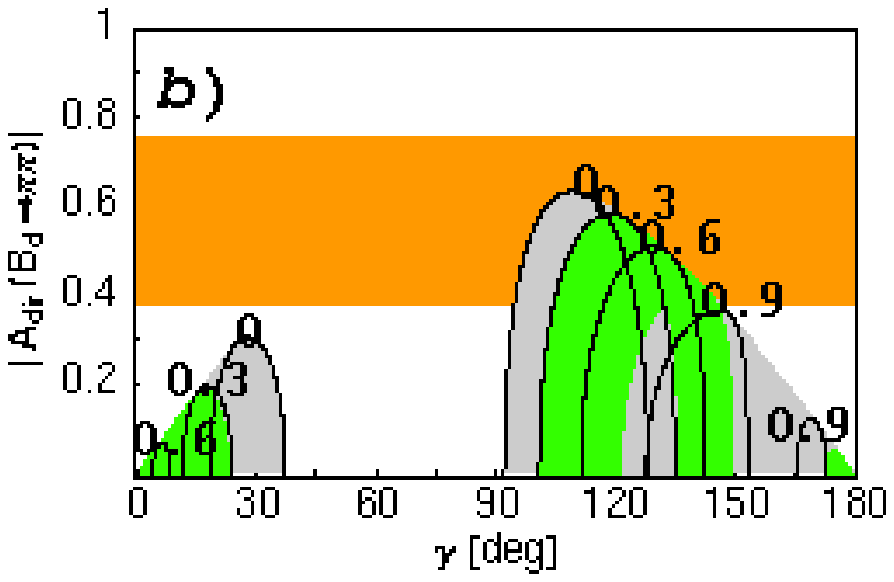}
$$
\vspace*{-0.2cm}
$$\hspace*{-1.cm}
\epsfysize=0.19\textheight
\epsfxsize=0.29\textheight
\epsffile{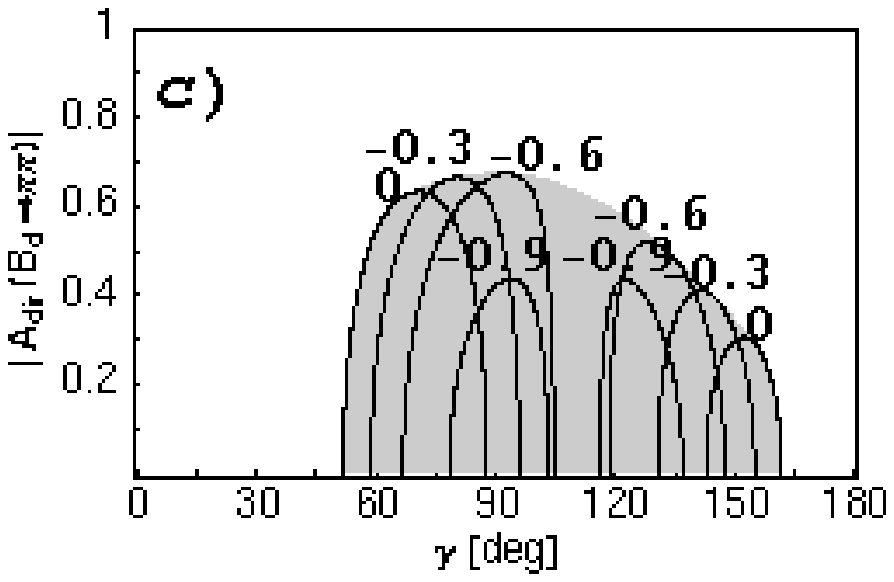} \hspace*{0.3cm}
\epsfysize=0.19\textheight
\epsfxsize=0.29\textheight
 \epsffile{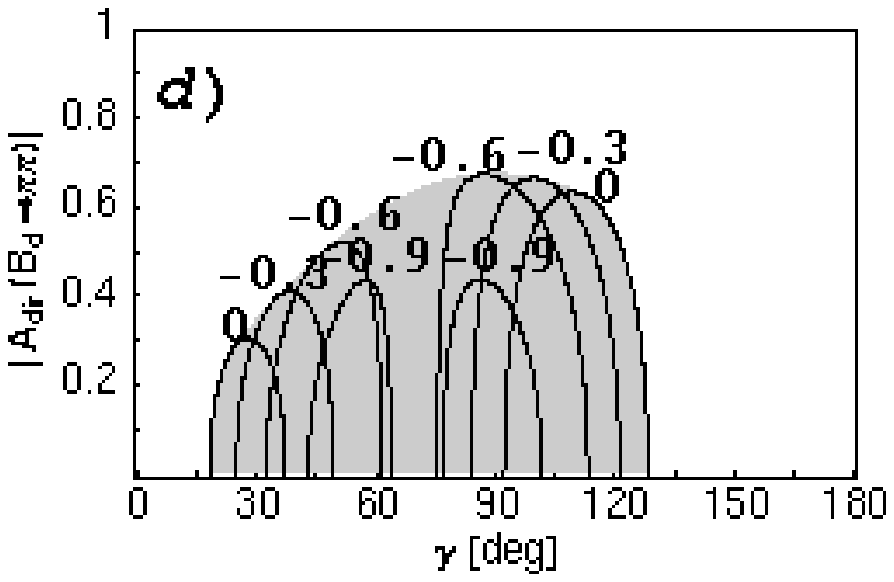}
$$
\caption[]{$|{\cal A}_{\rm CP}^{\rm dir}(B_d\to\pi^+\pi^-)|$ as a function
of $\gamma$ in the case of $H=7.5$ for various values of 
${\cal A}_{\rm CP}^{\rm mix}(B_d\to\pi^+\pi^-)$. 
In (a) and (b), $\phi_d=47^\circ$ and $\phi_d=133^\circ$ were chosen,
respectively. The shaded ``hills'' arise from a variation of 
${\cal A}_{\rm CP}^{\rm mix}(B_d\to\pi^+\pi^-)$ within $[0,+1]$. The
corresponding plots for negative  
${\cal A}_{\rm CP}^{\rm mix}(B_d\to\pi^+\pi^-)$ are shown in (c) and
(d) for $\phi_d=47^\circ$ and $\phi_d=133^\circ$, respectively. The 
bands arising from the experimental averages given in 
(\ref{Bpipi-CP-averages}) and (\ref{Bpipi-CP-averages2})
are also included.}\label{fig:gam-Add}
\end{figure}

In Fig.~\ref{fig:AdAmpipi}, we show the corresponding results for the two 
solutions of $\phi_d$ and for various values of $H$, as well as the contours 
arising for fixed values of $\gamma$ \cite{FlMa2}. We observe that the 
experimental averages, represented by the crosses, overlap nicely with the 
SM region for $\phi_d=47^\circ$, and point towards $\gamma\sim55^\circ$. 
In this case, not only $\gamma$ would be in accordance with the results 
of the CKM fits described in Section~\ref{sec:intro}, but also the 
$B^0_d$--$\overline{B^0_d}$ mixing phase $\phi_d$. On the other hand, for 
$\phi_d=133^\circ$, the experimental values favour $\gamma\sim125^\circ$, 
and have essentially no overlap with the SM region. Since a value of 
$\phi_d=133^\circ$ would require CP-violating new-physics contributions 
to $B^0_d$--$\overline{B^0_d}$ mixing, also the $\gamma$ range in 
(\ref{gamma-SM}) may no longer hold in this case, as it relies on a 
Standard-Model interpretation of the experimental information on 
$B^0_{d,s}$--$\overline{B^0_{d,s}}$ mixing. In particular, also values 
for $\gamma$ larger than $90^\circ$ could then in principle be accommodated. 
In order to put these observations on a more quantitative basis, we show in 
Fig.~\ref{fig:gam-Add} the dependence of 
$|{\cal A}_{\rm CP}^{\rm dir}(B_d\to\pi^+\pi^-)|$ on 
$\gamma$ for given values of ${\cal A}_{\rm CP}^{\rm mix}(B_d\to\pi^+\pi^-)$
\cite{FlMa2}. If we vary ${\cal A}_{\rm CP}^{\rm mix}(B_d\to\pi^+\pi^-)$ 
within its whole positive range $[0,+1]$, the shaded ``hills'' 
in Figs.~\ref{fig:gam-Add} (a) and (b) arise. In the case of 
$\phi_d=47^\circ$, which is in agreement with the CKM fits, we may 
conveniently accommodate the Standard-Model range (\ref{gamma-SM}). 
On the other hand, we obtain a gap around $\gamma\sim 60^\circ$ for 
$\phi_d=133^\circ$. Taking into account the experimental averages given 
in (\ref{Bpipi-CP-averages}) and (\ref{Bpipi-CP-averages2}), we obtain 
\begin{equation}\label{gam-res}
34^\circ\lsim\gamma\lsim75^\circ \, (\phi_d=47^\circ), \quad
105^\circ\lsim\gamma\lsim146^\circ \, (\phi_d=133^\circ).
\end{equation}
If we vary ${\cal A}_{\rm CP}^{\rm mix}(B_d\to\pi^+\pi^-)$ within its
whole negative range, both solutions for $\phi_d$ could accommodate 
(\ref{gamma-SM}), as can be seen in Figs.~\ref{fig:gam-Add} (c) and (d), 
so that the situation would not be as exciting as for a positive value of 
${\cal A}_{\rm CP}^{\rm mix}(B_d\to\pi^+\pi^-)$. In the future, the 
experimental uncertainties will be reduced considerably, i.e.\ the 
experimental bands in Fig.~\ref{fig:gam-Add} will become much more narrow, 
thereby providing significantly more stringent results for $\gamma$, 
as well as the hadronic parameters. For a detailed discussion of the 
corresponding theoretical uncertainties, as well as simplifications
that could be made through factorization, see \cite{FlMa2}.

\begin{figure}[t]
$$\hspace*{-1.cm}
\epsfysize=0.2\textheight
\epsfxsize=0.3\textheight
\epsffile{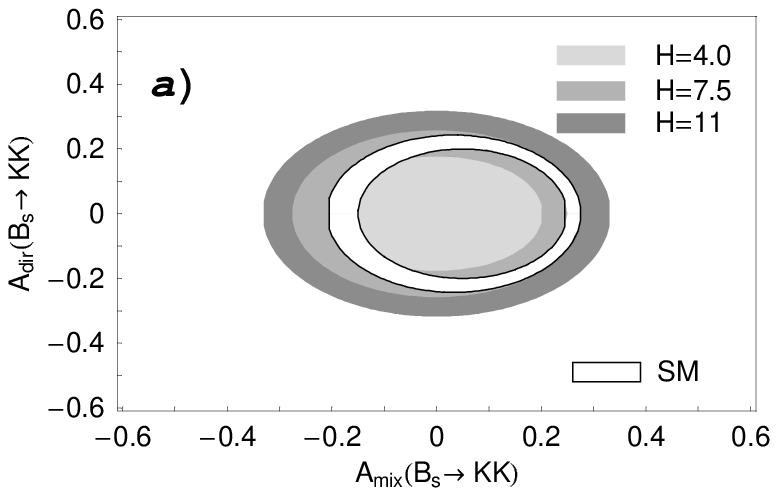} \hspace*{0.3cm}
\epsfysize=0.2\textheight
\epsfxsize=0.3\textheight
 \epsffile{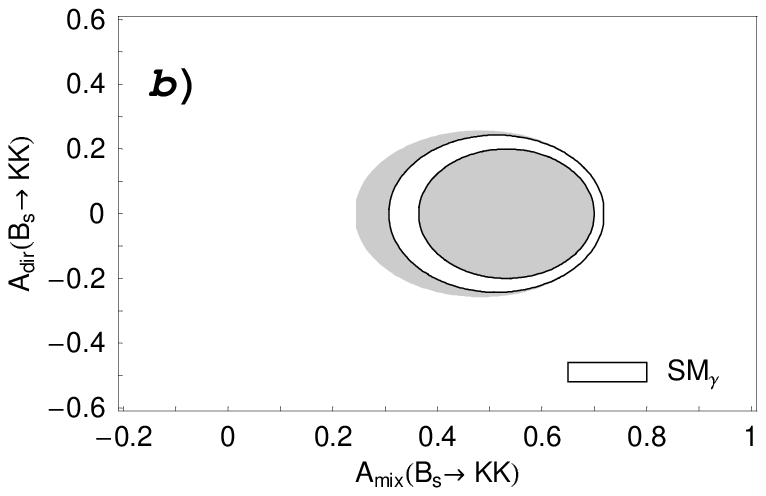}
$$
\caption[]{Allowed regions in the
${\cal A}_{\rm CP}^{\rm mix}(B_s\to K^+K^-)$--${\cal A}_{\rm CP}^{\rm
dir}(B_s\to K^+K^-)$ plane for (a) $\phi_s=0^\circ$ and various values of 
$H$, and (b) $\phi_s^{\rm NP}=30^\circ$ ($H=7.5$). The SM regions arise if 
$\gamma$ is restricted to (\ref{gamma-SM}).}\label{fig:Ams-Ads}
\end{figure}

In analogy to the analysis of the $B_d\to\pi^+\pi^-$ mode discussed 
above, we may also use $H$ to eliminate $d'$ in 
${\cal A}_{\rm CP}^{\rm dir}(B_s\to K^+K^-)$ and 
${\cal A}_{\rm CP}^{\rm mix}(B_s\to K^+K^-)$. If we then vary 
$\theta'$ and $\gamma$ within their physical ranges, i.e.\
$-180^\circ\leq \theta'\leq+180^\circ$ and $0^\circ\leq \gamma \leq180^\circ$,
we obtain an allowed region in the ${\cal A}_{\rm CP}^{\rm mix}(B_s\to 
K^+K^-)$--${\cal A}_{\rm CP}^{\rm dir}(B_s\to K^+K^-)$ plane \cite{FlMa2},
as shown in Fig.~\ref{fig:Ams-Ads}. There, also the impact of a
non-vanishing value of $\phi_s$, which may be due to new-physics contributions
to $B^0_s$--$\overline{B^0_s}$ mixing, is illustrated. If we constrain 
$\gamma$ to (\ref{gamma-SM}), even more restricted regions arise. The
allowed regions are remarkably stable with respect to variations of 
parameters characterizing $U$-spin-breaking effects \cite{FlMa2}, and 
represent a narrow target range for run II of the Tevatron and the 
experiments of the LHC era, in particular LHCb and BTeV. These
experiments will allow us to exploit the whole physics potential
of the $B_d\to\pi^+\pi^-$, $B_s\to K^+K^-$ system \cite{RF-BsKK}.

\section{Remarks on the ``Usual'' Rare $B$ Decays}\label{sec:rare}
Let us finally comment briefly on other ``rare'' $B$ decays, which
occur only at the one-loop level in the Standard Model, and involve 
$\overline{b}\to \overline{s}$ or $\overline{b}\to \overline{d}$ 
flavour-changing neutral-current transitions. Prominent examples are
the following decay modes:  $B\to K^\ast\gamma$, $B\to \rho\gamma$, 
$B\to K^\ast\mu^+\mu^-$ and $B_{s,d}\to \mu^+\mu^-$. The corresponding 
inclusive decays, for example $B\to X_s\gamma$, are also of particular
interest, suffering from smaller theoretical uncertainties. Within the 
Standard Model, these transitions exhibit small branching ratios at 
the $10^{-4}$--$10^{-10}$ level, do not -- apart from $B\to \rho\gamma$ -- 
show sizeable CP-violating effects, and depend on $|V_{ts}|$ or $|V_{td}|$. 
A measurement of these CKM factors through such decays would be 
complementary to the one from $B^0_{s,d}$--$\overline{B^0_{s,d}}$ mixing. 
Since rare $B$ decays are absent at the tree level in the Standard Model, 
they represent interesting probes to search for new physics. For detailed 
discussions of the many interesting aspects of rare $B$ decays, the reader 
is referred to the lecture given by Mannel at this school \cite{mannel}, 
and to the overview articles listed in \cite{isidori,rare}.

\section{Conclusions and Outlook}\label{sec:concl}
The phenomenology of the $B$ system is very rich and represents
an exciting field of research. Thanks to the efforts of the BaBar and 
Belle collaborations, CP violation could recently be established 
in the $B$ system with the help of the ``gold-plated'' mode 
$B_d\to J/\psi K_{\rm S}$, thereby opening a new era in the exploration 
of CP-violating phenomena. The world average $\sin2\beta=0.734\pm0.054$
agrees now well with the Standard Model, but leaves a twofold solution 
for $\phi_d$, given by $\phi_d=\left(47^{+5}_{-4}\right)^\circ \, \lor \,
\left(133^{+4}_{-5}\right)^\circ$. The former solution is in accordance 
with the picture of the Standard Model, whereas the latter would point 
towards CP-violating new-physics contributions to $B^0_d$--$\overline{B^0_d}$ 
mixing. As we have seen, it is an important issue to resolve this ambiguity 
directly.

The physics potential of the $B$ factories goes far beyond the
famous $B_d\to J/\psi K_{\rm S}$ decay, allowing us now to confront many 
more strategies to explore CP violation with data. Here the main goal is 
to overconstrain the unitarity triangle as much as possible, thereby 
performing a stringent test of the KM mechanism of CP violation. In 
this respect, important benchmark modes are given by $B\to\pi\pi$, 
$B\to\phi K$ and $B\to\pi K$ decays. First exciting data on these 
channels are already available from the $B$ factories, but do not yet 
allow us to draw definite conclusions. In the future, the picture 
should, however, improve significantly. 

Another important element in the testing of the Standard-Model 
description of CP violation is the $B_s$-meson system, which is not
accessible at the $e^+e^-$ $B$ factories operating at the $\Upsilon(4S)$
resonance, BaBar and Belle, but can be studied nicely at hadron collider 
experiments. Already, run II of the Tevatron is expected to provide interesting
results on $B_s$ physics, and should discover $B^0_s$--$\overline{B^0_s}$
mixing soon, which is an important ingredient for the ``standard'' analysis
of the unitarity triangle. Important $B_s$ decays are $B_s\to J/\psi \phi$, 
$B_s\to K^+K^-$ and $B_s\to D^\pm_s K^\mp$. Although the Tevatron will
provide first insights into these modes, they can only be fully exploited
at the experiments of the LHC era, in particular LHCb and BTeV. 

Apart from issues related to CP violation, several $B$-decay strategies 
allow also the determination of hadronic parameters, which can then be
compared with theoretical predictions and may help us to control the
corresponding hadronic uncertainties in a better way. Moreover, there 
are many other exciting aspects of $B$ physics, for instance studies of 
certain rare $B$ decays, which represent also sensitive probes for new 
physics. Hopefully, the future will bring many surprising results!

%
%
%
%
%

%

\end{document}